\documentclass[iop]{emulateapj}
\usepackage{natbib}
\usepackage{graphicx}
\usepackage{times}
\usepackage{xspace} 
\usepackage{booktabs} 
\usepackage{amsmath}
\usepackage{subfigure}
\usepackage{float}
\usepackage{afterpage}

\slugcomment{\it Accepted Version --- 2017 Feb}
\shorttitle{Most massive AGNs}
\shortauthors{Jun et al.}

\def\deg{\ifmmode {^{\circ}}\else {$^\circ$}\fi}
\def\kms{\ifmmode {\rm\,km\,s^{-1}}\else
    ${\rm\,km\,s^{-1}}$\fi}
\def\ergcm2s{\ifmmode {\rm\,erg\,cm^{-2}\,s^{-1}}\else
    ${\rm\,erg\,cm^{-2}\,s^{-1}}$\fi}
\def\ergAcm2s{\ifmmode {\rm\,erg\,cm^{-2}\,s^{-1}\,\AA^{-1}}\else
    ${\rm\,erg\,cm^{-2}\,s^{-1}\,\AA^{-1}}$\fi}
\def\ergs{\ifmmode {\rm\,erg\,s^{-1}}\else
    ${\rm\,erg\,s^{-1}}$\fi}
\def\kmsMpc{\ifmmode {\rm\,km\,s^{-1}\,Mpc^{-1}}\else
    ${\rm\,km\,s^{-1}\,Mpc^{-1}}$\fi}

\begin{document}
\title{The Most Massive Active Galactic Nuclei at $1\lesssim z \lesssim 2$}
\author{Hyunsung D. Jun\altaffilmark{1,2,3}, Myungshin Im\altaffilmark{3,4}, Dohyeong Kim\altaffilmark{3}, and Daniel Stern\altaffilmark{1}}
\altaffiltext{1}{Jet Propulsion Laboratory, California Institute of Technology, 4800 Oak Grove Drive, Pasadena, CA 91109, USA; hyunsung.jun@jpl.nasa.gov}
\altaffiltext{2}{NASA Postdoctoral Program Fellow.}
\altaffiltext{3}{Center for the Exploration of the Origin of the Universe (CEOU), Astronomy Program, Department of Physics 
and Astronomy, Seoul National University, Seoul 151-742, Korea}
\altaffiltext{4}{Author to whom correspondence should be addressed; mim@astro.snu.ac.kr}

\begin{abstract}
We obtained near-infrared spectra of 26 SDSS quasars at $0.7<z<2.5$ with reported rest-frame ultraviolet $M_{\rm BH} \sim 10^{10}M_{\odot}$ to critically examine the systematic effects involved with their mass estimations. We find that AGNs heavier than $10^{10}M_{\odot}$ often display double-peaked H$\alpha$ emission, extremely broad \ion{Fe}{2} complex emission around \ion{Mg}{2}, and highly blueshifted and broadened \ion{C}{4} emission. The weight of this evidence, combined with previous studies, cautions against the use of $M_{\rm BH}$ values based on any emission line with a width over 8000\kms. Also, the $M_{\rm BH}$ estimations are not positively biased along the presence of ionized narrow line outflows, anisotropic radiation, or the use of line FWHM instead of $\sigma$ for our sample, and unbiased with variability, scatter in broad line equivalent width, or obscuration for general type-1 quasars. Removing the systematically uncertain $M_{\rm BH}$ values, $\sim10^{10}M_{\odot}$ BHs in $1\lesssim z \lesssim 2$ AGNs can still be explained by anisotropic motion of the broad line region from $\sim10^{9.5}M_{\odot}$ BHs, although current observations support they are intrinsically most massive, and overmassive to the host's bulge mass.
\end{abstract}


\section{Introduction}
Since the identification of supermassive black holes (BHs) at the center of galaxies, their typical mass ($M_{\rm BH}$) values have been measured in the $10^{6-9} M_{\odot}$ range (e.g., \citealt{Kor95}). The more recent discovery of $\sim$10$^{10} M_{\odot}$ BHs \citep{McC11} in quiescent galaxies further extended the massive limit, pushing the previous $\sim$10$^{9}M_{\odot}$ boundary to heavier regimes. These extremely massive black holes ($>$10$^{9.5} M_{\odot}$, hereafter EMBHs) give constraints to how massive a BH can grow through accretion within the inner galaxy (e.g., \citealt{Ina16}; \citealt{Kin16}). Also, the EMBHs are thought to reside in $\sim$10$^{12} M_{\odot}$ giant elliptical host galaxies lying on the present day $M_{\rm BH}$--$\sigma_{*}$ relation (e.g., \citealt{Fer00}; \citealt{Geb00}), though we note there are exceptions to the expectation that BH growth closely follows that of the host (e.g., \citealt{van12}; \citealt{Set14}; \citealt{Wal15}). 

Direct evidence for the existence of 10$^{10} M_{\odot}$ BHs was initially reported in a handful of nearby quiescent galaxies (e.g., NGC 3842 and NGC 4889, \citealt{McC11}; NGC 1277, \citealt{van12}), albeit with the validity of some of the measurements being questioned \citep{Ems13}. Even if we consider that the measured values are acceptable, the EMBHs tend to lie above the $M_{\rm BH}$--$\sigma_{*}$, or more frequently, the $M_{\rm BH}$--$L_{\rm{bulge}}$ relations extrapolated from lower mass BHs (e.g., \citealt{Gul09}; \citealt{Kor13}), by up to an order of magnitude (see also, \citealt{Sav16} on measurement issues for $L_{\rm{bulge}}$). To explain the high mass of EMBHs with respect to their host galaxies, \citet{Vol13} suggest that EMBHs are formed through frequent dry mergers. However, this does not solve the problem entirely because such mergers might not necessarily induce strong active galactic nucleus (AGN) activity in EMBHs found up to at least $z=5$ (e.g., \citealt{Jun15}, hereafter J15; \citealt{Wu15}).

These observational and theoretical considerations lead to the natural question if the estimates of $M_{\rm BH}\sim10^{10} M_{\odot}$ in AGNs are reliable. The $M_{\rm BH}$ estimators applied to high redshift quasar spectra have mainly relied on UV-based spectral features that are secondarily calibrated to Hydrogen Balmer line based estimators. One possibility is that the UV-based $M_{\rm BH}$ values are overestimated somehow. Independent $M_{\rm BH}$ estimates from Balmer-line based estimators would enhance the reliability of 10$^{10} M_{\odot}$ BHs in distant quasars. Unfortunately, direct comparison between the UV-based versus Balmer-based $M_{\rm BH}$ estimates has been scarce for EMBHs. Previous studies have been largely limited to $M_{\rm BH}<$10$^{9.5} M_{\odot}$ (e.g., \citealt{Net07}; \citealt{Sha07}; \citealt{Die09}; \citealt{Gre10}; \citealt{Ho12}; \citealt{Par13}), and such studies are controversial regarding the scatter between \ion{C}{4} and Balmer-based estimators while consistent on the agreement betwen \ion{Mg}{2} and Balmer-based $M_{\rm BH}$ values. The most extensive study in this respect was done by Shen et al. (2012, hereafter S12) where the sample includes dozens of EMBHs, though with few BHs above 10$^{10} M_{\odot}$.

In order to understand the mass growth of EMBHs, accurate $M_{\rm BH}$ measurements over a range of redshifts is vital. In the distant universe, however, direct dynamical measurement of $M_{\rm BH}$ becomes difficult for quiescent galaxies, as it is hard to resolve the gravitational sphere of influence from the BH. Instead, broad line gas kinematics are used to estimate the $M_{\rm BH}$ of AGNs, where this method gives more uncertain results. $M_{\rm BH}$ measurements for AGNs are based on the reverberation mapping technique (\citealt{Bla82}; \citealt{Pet93}) which measures the time delay of the broad line emission to the incident continuum and thus estimates the size of the broad line region ($R_{\rm BLR}$). The $R_{\rm BLR}$ values for H$\beta$ are further calibrated by the radius--luminosity relationship ($R_{\rm BLR }$--$L$ relation, \citealt{Kas00}; \citealt{Ben06}), which allows the estimation of $M_{\rm BH}$ from a single-epoch measurement of optical continuum/line luminosity and broad line width. 

High redshift AGNs have their rest-frame optical (rest-optical) emission redshifted to the infrared, and their rest-frame ultraviolet (rest-UV) emission redshifted into the optical. The single-epoch mass estimators are thus secondarily calibrated in the UV assuming that the UV continuum luminosity and broad emission line (\ion{C}{4} or \ion{Mg}{2}) widths follow the optical $R_{\rm BLR }$--$L$ relation and optical line widths respectively, as a linear or power-law relation. Ongoing studies tentatively find that the slope of the \ion{C}{4} $R_{\rm BLR }$--$L$ relation follows that of the optical relation, and extends up to the most luminous quasars (\citealt{Kas07}; \citealt{Slu11}; \citealt{Che12}). However, the UV continuum luminosities and line widths are not tightly correlated with the optical quantities, introducing an intrinsic scatter of $\sim$\,0.4\,dex when comparing \ion{C}{4} and Balmer $M_{\rm BH}$ measurements (e.g., J15).

In addition to the issues regarding the reliability of the rest-UV $M_{\rm BH}$ measurements, the rest-optical $M_{\rm BH}$ measurement from single-epoch spectroscopy itself has limitations on its accuracy due to the systematic uncertainties in deriving the mass equation. Bearing in mind that the single-epoch $M_{\rm BH}$ values have sizable errors from the poorly constrained constant (virial factor, hereafter $f$--factor) in the mass equation (0.3--0.4\,dex systematic uncertainty, e.g., \citealt{Kor13}; \citealt{McC13}) and the $R_{\rm BLR }$--$L$ relation (0.1--0.2\,dex intrinsic scatter, \citealt{Ben13}), it is possible that the $M_{\rm BH}$ values of the most extreme AGNs could have been biased to high values if they were selected to have outlying $f$-factors or $R_{\rm BLR}$ values with respect to the calibrations. Indeed, theoretical and technical issues that could positively bias the $M_{\rm BH}$ measurements have been reported, from accretion disk modeling (\citealt{Lao11}; \citealt{Wan14}) or profile fit methodology (\citealt{Pet04}; \citealt{Col06}). Furthermore, potential limitations of automated spectral fitting of a large sample of spectra failing to model unusual spectral features (e.g., \citealt{She08}) or using single-epoch spectroscopy to derive representative AGN properties should be carefully checked, especially at extreme mass values.

In this paper, we present rest-optical spectra of 26 quasars at 0.7\,$<$\,$z$\,$<$\,2.5 with UV-based $\sim$10$^{10} M_{\odot}$ $M_{\rm BH}$ measurements, obtained with the NASA Infrared Telescope Facility (IRTF). We aim to double check the consistency of the massive end UV-optical $M_{\rm BH}$ estimates, and examine if the measurements could be systematically biased to unusually high masses from spectral features and during application of the mass estimator. We describe the sample selection and data acquisition of extremely massive AGNs (section 2), the spectral analysis in determining $M_{\rm BH}$ (section 3), the results (section 4) and implications on the measured $M_{\rm BH}$ values (section 5). Throughout, we adopt a flat $\Lambda$CDM cosmology with $H_{0}=\mathrm{70\,km\,s^{-1}\,Mpc^{-1}}$, $\Omega_{m}=0.3$, and $\Omega_{\Lambda}=0.7$ (e.g., \citealt{Im97}).

\section{Data}
\subsection{Sample description and observations}
We selected the extremely massive AGN sample from the Sloan Digital Sky Survey (SDSS) type-1 quasar catalog (DR7, \citealt{Sch10}). The spectral fitting results and the $M_{\rm BH}$ estimates in \citet{She11} were adopted for target selection, using the H$\beta$ line at $z<0.8$, \ion{Mg}{2} at $0.8<z<2.0$, and \ion{C}{4} at $z>2.0$. We identified the sample by applying the following selection criteria:\\
\begin{itemize}
\item Mass selection of $M_{\rm BH} \geq 5\times10^{9}M_{\odot}$
\item Redshift cut of $0.7<z<2.5$ to place the broad H$\alpha$ and H$\beta$ lines within the near-infrared (NIR) spectroscopic windows, excluding redshifts where both H$\alpha$ and H$\beta$ are close to NIR telluric absorption ($1.1<z<1.3$ and $1.6<z<2.2$)
\item Continuum signal-to-noise ratio (S/N) cut of 20 or higher from the SDSS spectra for good line width and flux measurements, $H$-band magnitude $<$\,17.8 AB mag, bright enough for IRTF observations
\item Removal of objects flagged or visually inspected to show double-peaked H$\beta$ lines, severely absorbed \ion{Mg}{2} or \ion{C}{4}, and obvious mismatch between the model and the spectrum
\end{itemize}
yielding 1254 quasars sufficing the $M_{\rm BH}$ and redshift cut, and 261 objects with further sensitivity limits and flags, among which we arbitrarily selected 26 objects spanning the optical continuum luminosities\footnote{Throughout this paper we use subscript numbers on the monochromatic luminosity to indicate its wavelength, such as $L_{5100}=L(5100\rm \AA)$.} of $L_{5100}=10^{45.7-47.2}$\ergs, for IRTF observations. Figure 1 shows the redshift--$M_{\rm BH}$ distribution of our sample.

\begin{figure}
\centering
\includegraphics[scale=.95]{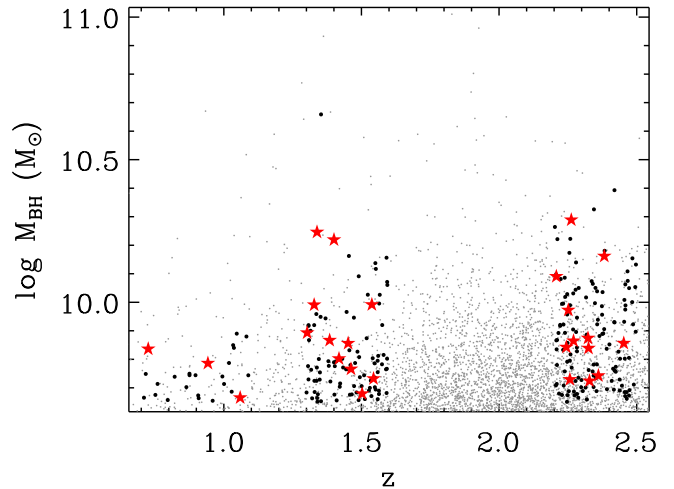}
\caption{The distribution of the massive end $M_{\rm BH}$ values at intermediate redshifts, with the measurements from Shen et al. (2011, gray dots). Those within our redshift and sensitivity cuts, flagged or visually inspected for spurious line profiles or fitting, are displayed (black circles), among which we picked 26 for IRTF follow-up spectroscopy (red stars).}
\end{figure}

We used the SpeX instrument \citep{Ray03} on IRTF to obtain the NIR spectra of the targets. The 0.8--2.4\,$\mu$m cross-dispersed mode (SXD) was chosen, with the slit width of 0.8$\arcsec$ or 1.6$\arcsec$ depending on the seeing conditions. This yields a spectral resolution of $R=750$ or 375 throughout the observed wavelengths, which is tuned for sensitivity over resolution for the broad AGN emission lines and continuum features. The exposure times for the targets were aimed to give a continuum S/N per resolution element of at least 10 around H$\beta$, and 5 around H$\alpha$. We carefully checked each field to avoid neighbor source contamination when nodding the spectrum along the 15$\arcsec$ slit length. The observations were performed during three full and three half nights in 2011 December, 2012 February/December along with another program, summing to a total on-source integration of 9.7 hours for all the targets (0.1--0.6 hours per target). The weather conditions were overall photometric but atmospheric seeing varied from 0.6--2$\arcsec$. We nodded the spectra in AB mode with a 90--180s frame time for good dark and sky subtraction, and observed A0 V type standard stars nearby the target for telluric absorption correction \citep{Vac03} and flux calibration. Also, a set of flat-field and argon arc wavelength calibration data were taken. We summarize the IRTF observations in Table 1. 

In addition to the NIR spectroscopy, we compiled the optical spectra of quasars from the SDSS database (DR12 including both the SDSS-I/SDSS-II and the SDSS-III BOSS data, \citealt{Ala15}) in order to calculate the $M_{\rm BH}$ values from \ion{C}{4} and \ion{Mg}{2} lines and to compare them with masses from hydrogen Balmer lines or previous rest-UV measurements. Also, broad-band photometric data from GALEX GR7, SDSS DR12, 2MASS PSC, UKIDSS DR10, and WISE AllWISE releases (\citealt{Mar05}; \citealt{Skr06}; \citealt{Law07}; \citealt{Wri10}; \citealt{Ala15}) were collected to supplement the spectra with monochromatic continuum luminosities. The latest spectra (SDSS-III BOSS over SDSS-I/SDSS-II) and photometry (UKIDSS over 2MASS) were used when the target had overlapping data, while multiple spectra from the same instrument were averaged. 

\begin{deluxetable}{cccccc}
\tablecolumns{4}
\tablecaption{Summary of IRTF observations}
\tablewidth{0.45\textwidth}
\tablehead{
\colhead{Name} & \colhead{Coordinates} & \colhead{$z$} & \colhead{$H$} & \colhead{$t_{\rm exp }$} & \colhead{$R$}}
\startdata
J0102+00 & J010205.89+001157.0 & 0.727 & 16.85 & 30 & 750\\
J1010+05 & J100943.56+052953.9 & 0.944 & 17.07 &  9 & 375\\
J0748+22 & J074815.44+220059.5 & 1.060 & 16.10 & 24 & 750\\
J0840+23 & J083937.85+223940.7 & 1.312 & 16.25 & 18 & 750\\
J1057+31 & J105705.16+311907.9 & 1.329 & 17.30 & 24 & 750\\
J0203+13 & J020256.11+124928.0 & 1.352 & 17.49 & 30 & 750\\
J0319--07 & J031926.24--072808.8 & 1.391 & 16.78 & 30 & 375\\
J1053+34 & J105250.06+335504.9 & 1.414 & 16.40 & 12 & 750\\
J1035+45 & J103453.06+445723.2 & 1.424 & 15.25 &  9 & 750\\
J1055+28 & J105440.84+273306.4 & 1.453 & 17.12 & 18 & 750\\
J0146--10 & J014542.78--100807.7 & 1.465 & 16.76 & 24 & 750\\
J0400--07 & J040022.40--064928.6 & 1.516 & 16.58 & 30 & 375\\
J0855+05 & J085515.59+045232.8 & 1.541 & 17.08 & 18 & 750\\
J0741+32 & J074043.47+314201.2 & 1.546 & 17.70 & 24 & 750\\
J1522+52 & J152156.48+520238.6 & 2.221 & 15.44 &  6 & 750\\
J1339+11 & J133928.39+105503.2 & 2.250 & 17.18 & 12 & 750\\
J0905+24 & J090444.34+233354.1 & 2.258 & 16.62 & 30 & 750\\
J0257+00 & J025644.69+001246.0 & 2.264 & 17.78 & 30 & 750\\
J2123--01 & J212329.47--005052.9 & 2.282 & 15.90 & 12 & 750\\
J0052+01 & J005202.41+010129.2 & 2.283 & 17.07 & 18 & 750\\
J2112+00 & J211157.78+002457.5 & 2.335 & 17.58 & 36 & 375\\
J1027+30 & J102648.16+295410.9 & 2.349 & 16.92 & 24 & 750\\
J0651+38 & J065101.23+380759.6 & 2.355 & 17.75 & 30 & 375\\
J1036+11 & J103546.03+110546.5 & 2.368 & 16.45 & 18 & 750\\
J0752+43 & J075158.65+424522.9 & 2.466 & 17.64 & 24 & 375\\
J0946+28 & J094602.31+274407.1 & 2.476 & 16.66 & 18 & 750
\enddata
\tablecomments{$z$ is the redshift of the H$\alpha$ line (section 3), $H$ is the $H$-band AB magnitude, $t_{\rm exp}$ is the total
exposure time in minutes, and $R$ is the spectral resolution.} 
\end{deluxetable} 

\begin{figure*}
\centering
\includegraphics[scale=.665]{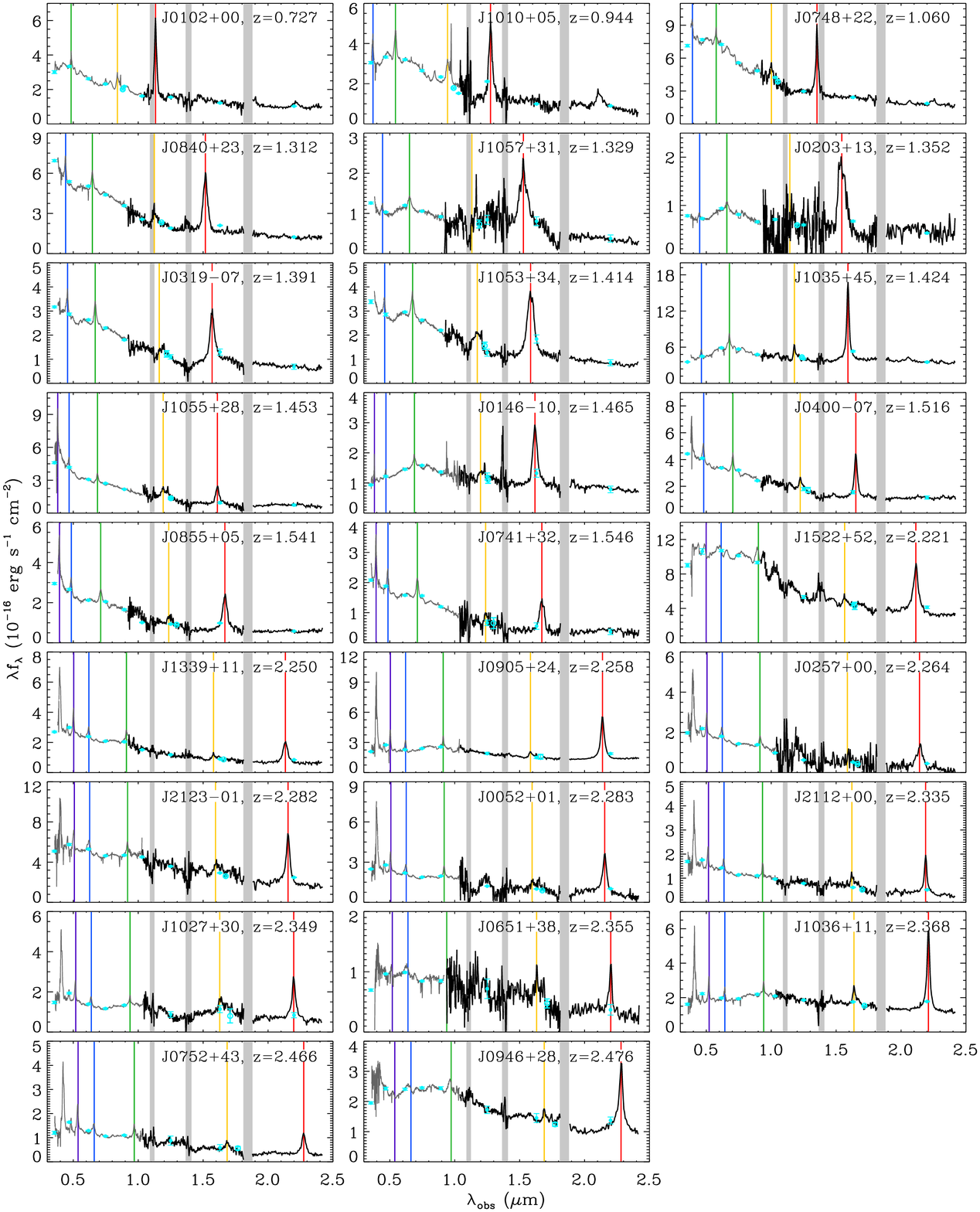}
\caption{The reduced observed-frame SDSS (gray) and IRTF (black) spectra of the sample, binned to $R\sim400$ for display purposes. Gray shaded regions are wavelengths with strong NIR telluric absorption. The photometric data points from SDSS, 2MASS or UKIDSS, are overplotted (filled cyan circles), together with the rest-frame 5100\,\AA\ continuum fluxes (open cyan circles). The H$\alpha$, H$\beta$, \ion{Mg}{2}, \ion{C}{3}], and \ion{C}{4} emission lines are marked in thin red, yellow, green, blue, and purple lines, respectively.}
\end{figure*}

\subsection{Data reduction}
We reduced the IRTF spectra using the IDL-based package Spextool (version 3.4, \citealt{Cus04}). It involves pre-processing (linearity, flat correction), spectral extraction, wavelength and flux calibration, combining multiple spectral frames, telluric correction, order merging, and spectrum cleaning. The standard package configuration was adopted, with a seeing dependent, 0.7--1.2$\arcsec$ Gaussian spatial extraction radius set equal for the target and standard star spectra. We found up to a $\sim$10--20\% level of flux difference at overlaps between different orders of the cross-dispersed spectra, which were leveled using the Spextool package. Moreover, we checked the accuracy of standard star flux calibration by convolving each flux-calibrated IRTF spectrum by the broad SDSS/UKIRT filter response curves, and comparing the spectroscopic flux to that from the photometry. Overall, we find the mean and rms scatter of the spectroscopic to photometric flux ratio, to be $0.97\pm0.36$ when averaged over the set of $griz$ filters (not taking into account the $u$-band because the shortest wavelength cutoff is different for the photometry and the spectroscopy), and $0.91\pm0.35$ for the $YJHK$ filters. In order to reduce the scatter between the spectral and photometric fluxes, we linearly interpolated the flux ratios and gave multiplicative corrections to the SDSS and IRTF spectra, where the mean and rms scatter of the spectroscopic to photometric flux ratio change to $1.00\pm0.00$ and $1.01\pm0.07$ for the $griz$ and $YJHK$ bands, respectively. 

We plot the flux calibrated spectra together with the photometric data points in Figure 2. The SDSS and IRTF spectra meet fairly well at their boundaries though the short wavelength ($\sim$\,1\,$\mu$m) IRTF data are noisy due to some of the data taken in bright lunar phases and the weaker sensitivity of higher order spectra. The average and 1-$\sigma$ scatter of the continuum S/N\footnote{Throughout this paper we measure the S/N per wavelength element $\Delta \lambda=\lambda/750$ unless quoting the numbers from references.} are $89 \pm 30$ and $15\pm11$ for the SDSS and IRTF spectra, suitable for measuring $M_{\rm BH}$ for most of the targets except for some from the H$\beta$ line. We applied Galactic extinction corrections assuming the total-to-selective extinction ratio of $R_{V}=3.1$ and the $E(B-V)$ values from \citet{Bon00} that revised the values in the \citet{Sch98} extinction map.

\section{Analysis}
In order to estimate the single-epoch $M_{\rm BH}$ values from H$\beta$, H$\alpha$, \ion{Mg}{2}, \ion{C}{3}], and \ion{C}{4} lines, we fit the broad line regions from the joint SDSS/IRTF spectra\footnote{Throughout, we used the IDL-based package MPFIT \citep{Mar09} for all least-squares fitting unless stated otherwise.}. We start from the rest-frame 4200--5600\,\AA\ fit around H$\beta$ for which we used a power-law for the continuum, broad \ion{Fe}{2} component, and broad and narrow Gaussian components for the line. After fitting the power-law continuum determined by 4100--4300 and 5500--5700\,\AA\ windows and subtracting it from the spectrum, we determined the width (FWHM\,=\,900--20,000\,\kms) and height of the \ion{Fe}{2} complex using the \citet{Bor92} template while iteratively updating the continuum. We utilized the 4450--4650 and 5150--5350\,\AA\ regions to derive the height and the full fitting range to obtain the width of the \ion{Fe}{2}, through least chi-squares fit to the continuum subtracted spectrum. The H$\beta$ emission was fit by a single narrow (full width at half maximum, or FWHM\,$<$\,1000\,\kms\ hereafter\footnote{All the line widths mentioned in this paper are corrected for instrumental resolution, e.g., 400\,\kms\ for $R=750$ and 800\,\kms\ for $R=375$.}) Gaussian and double broad (FWHM\,=\,2000--15,000\,\kms\ hereafter) Gaussian components, double narrow Gaussians for each [\ion{O}{3}]$\lambda$4957, 5007 doublet, and H$\gamma$ by a single narrow and single broad Gaussian. When some of the [\ion{O}{3}] profiles were broader than their limit, the FWHM limit was relaxed to FWHM\,$<$\,2000\,\kms. To obtain a better quality fit, we used a common redshift for the narrow H$\beta$, one of the double narrow [\ion{O}{3}]$\lambda$4957 and one of the double narrow [\ion{O}{3}]$\lambda$5007, one of the double broad H$\beta$, and also the broad and narrow H$\gamma$, while leaving the rest of the components' redshift free. In addition, the centers of each Gaussian component were constrained to lie within 1000\,\kms\ of the H$\alpha$ redshift. We masked out or slightly modified the fitting range to exclude noisy regions that yielded a poor initial fit, and removed the fits where S/N\,$<$\,5. We obtained the monochromatic luminosity $L_{5100}$ and its error from the best fit model to the spectra with S/N\,$>$\,5. At S/N\,$<$\,5 we fit the photometric magnitudes at rest-frame 3000--10,000\,\AA\ by a power-law continuum, correcting for the H$\alpha$ line contribution using the $JHK$ filter response curves and the best fit model to the H$\alpha$ spectra, to obtain $L_{5100}$ depicted in Figure 2.

Next, we fit the rest-frame 6000--7100\,\AA\ region including the H$\alpha$ emission. We fixed the height and width of the \ion{Fe}{2} complex from the H$\beta$ region (or the \ion{Mg}{2} region when S/N$<$5) since they are weaker and harder to constrain around the H$\alpha$ emission. A power-law continuum, single narrow Gaussian and double broad Gaussians for the H$\alpha$, single narrow Gaussian for each [\ion{O}{1}]$\lambda$6300, 6364, [\ion{N}{2}]$\lambda$6548, 6583, and [\ion{S}{2}]$\lambda$6716, 6731 doublet, were simultaneously fitted to the \ion{Fe}{2} subtracted spectrum. We used the width of the [\ion{O}{3}]$\lambda$5007 from the H$\beta$ fit to fix the width of the crowded assembly of narrow [\ion{O}{1}] doublet, [\ion{N}{2}] doublet, H$\alpha$ singlet, and [\ion{S}{2}] doublet emission. When the [\ion{O}{3}] width was not reliably measured due to poor resolution/sensitivity, we used the mean FWHM of the narrow H$\alpha$, 400\,\kms, out of $z$\,$<$\,0.37, S/N\,$>$\,20 SDSS quasars from \citet{She11}. The relative strengths of the narrow [\ion{O}{1}], [\ion{N}{2}], and [\ion{S}{2}] lines were fixed to the values from \citet{Van01}. The centers of the narrow lines and one of the double broad H$\alpha$ components were tied to the same redshift. Also, the centers of every component were restricted to within 1000\,\kms\ of the H$\alpha$ redshift, which was determined by the peak of the broad H$\alpha$ model profile.

We notice that some objects show 
double-peaked features on top of the smooth, broad H$\alpha$, exhibiting relatively wide line widths (FWHM\,$\gtrsim$\,8000\,\kms) when fitted altogether (section 4.2.1). They resemble the disk emitters explained by a rotating accretion disk source pronounced in a small fraction of H$\alpha$ spectra, where the broad line luminosity or width using the full profile can overestimate the contribution from the BLR (e.g., Figure 1 in \citealt{Che89}; Figure 4 in \citealt{Era94}). In order to quantify which sources are likely disk emitters, we modeled again the broad H$\alpha$ spectra using a three component profile: one signal broad Gaussian, assumed to be the profile from the random motion of the BLR, and double broad Gaussians centered blue- and redwards from the H$\alpha$ redshift by more than 2000\,\kms\ which we assume to be the rotating disk emitter components. We define an emission line to be double-peaked when non-zero blue- and redshifted broad Gaussian components make up more than half of the total broad H$\alpha$ luminosity, the triple component model is favored over the single or double component fit with a statistically smaller reduced chi-square (a F-distribution probability of over 0.99), and the disk emitter components are clearly detached in order to distinguish from a simply wide and non-Gaussian BLR model. The criteria give seven classified double-peaked emitters\footnote{We note that our criteria indicate but do not verify with highly sensitive spectra or sophistcated modeling that these sources are disk emitters, thus we use the term double-peaked emitter throughout. These H$\alpha$ based double-peaked emitters are likely to be similar in their properties to the H$\beta$ based double-peaked emitters that we did not include in our sample (section 2.1). However, we kept them in our analysis to see how they affect $M_{\rm BH}$ estimates.}.

Third, we fit the rest-frame 2200--3100\,\AA\ region surrounding the \ion{Mg}{2} emission. We used the \ion{Fe}{2} template from \citet{Tsu06} because it provides data closer to the center of the \ion{Mg}{2} emssion than \citet{Ves01}. Following a methodology similar to the H$\beta$ fitting, we iteratively subtracted the continuum and \ion{Fe}{2} complex before fitting the \ion{Mg}{2}, determined through 2150--2250 and 3050--3150\,\AA\ windows for the continuum, 2150--2410, 2460--2700, 2900--3150\,\AA\ windows for the \ion{Fe}{2} height, and the full fitting range for the \ion{Fe}{2} width. Afterward, we fitted the \ion{Mg}{2} emission with double broad Gaussians and the [\ion{Ne}{4}]/\ion{Fe}{3} near 2420--2440\,\AA\ with a single broad Gaussian. We required the centers of the \ion{Mg}{2} and [\ion{Ne}{4}]/\ion{Fe}{3} model components to lie within 1000\,\kms\ of the H$\alpha$ redshift, with exceptions for J0946+28 and J1522+52 where the \ion{Mg}{2} centers are blueshifted by more and relaxed to lie between $-$3000 and 1000\,\kms. The \ion{Mg}{2} spectra showing absorption features, discontinuity between the SDSS/IRTF spectra, or imperfect \ion{Fe}{2} subtraction, were masked or adjusted in fitting range. We derived $L_{3000}$ and its error from the best fit model to the \ion{Mg}{2} region.

\begin{figure*}
\includegraphics[scale=0.99]{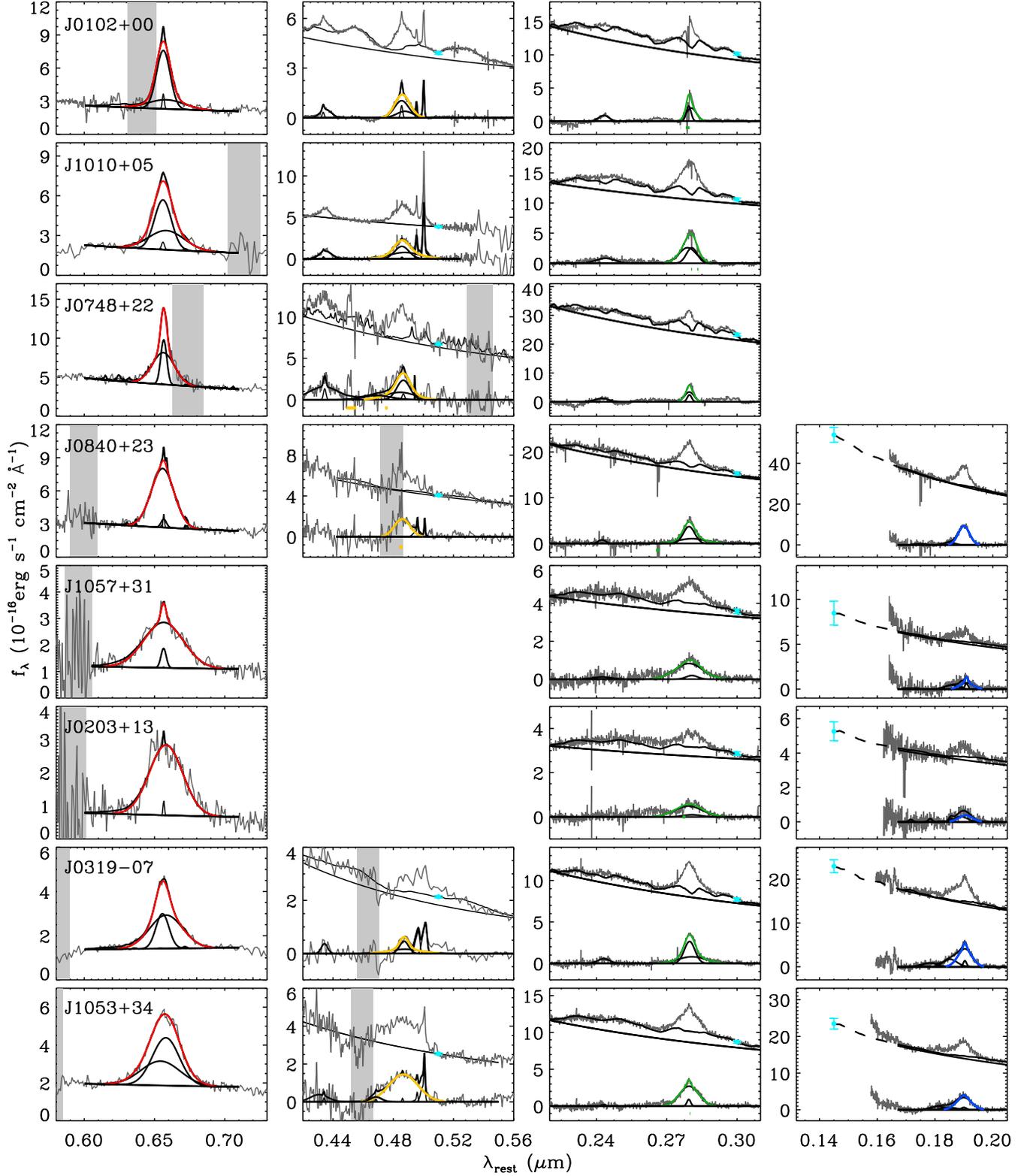}
\caption{The spectal fitting of H$\alpha$, H$\beta$, \ion{Mg}{2}, and \ion{C}{4} or \ion{C}{3}] line regions (from left to right). Five spectra with S/N$<$5 near the H$\beta$, one with a large discontinuity between the SDSS/IRTF data near the \ion{Mg}{2}, and three without \ion{C}{4} or \ion{C}{3}] coverage, are omitted. On top of the resolution matched spectra are the model narrow lines (thin black), broad lines (thick black lines for individual broad components, colored for the sum of the broad compoments used to derive $M_{\rm BH}$), the continuum (black line in the upper spectrum), the \ion{Fe}{2} complex (black curve on top of the continuum), and the sum of the total line components (black line in the lower spectrum). Masked regions are highlighted below the lower spectrum (thick lines colored identical to the broad emission), and the monochromatic luminosities, $L_{5100}$, $L_{3000}$, $L_{1350}$/$L_{1450}$ on the upper spectrum (cyan dots). Extrapolations to the $L_{1350}$/$L_{1450}$ are shown (black dashed line).
}
\end{figure*}
\begin{figure*}
\includegraphics[scale=0.99]{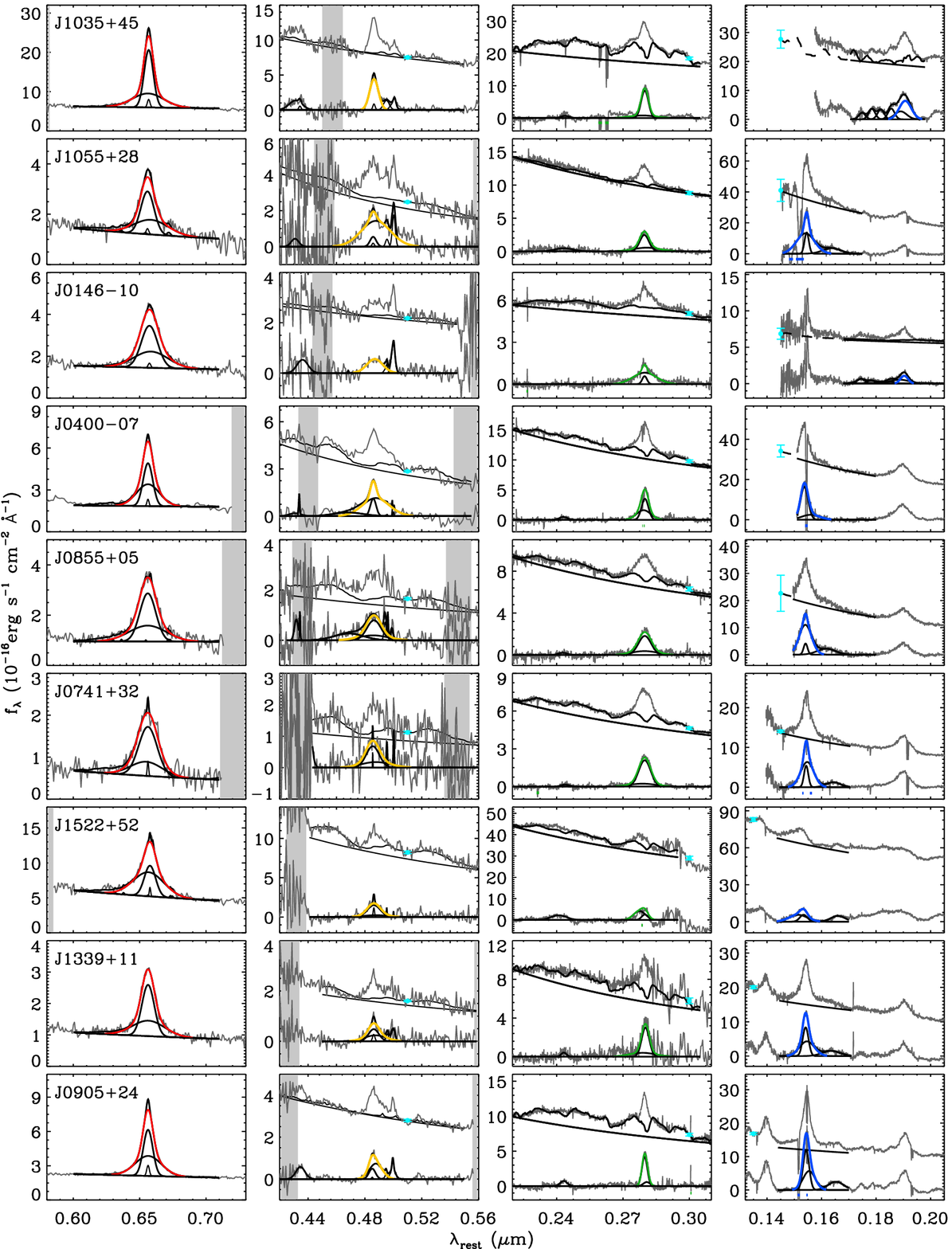}
\end{figure*}
\begin{figure*}
\includegraphics[scale=0.99]{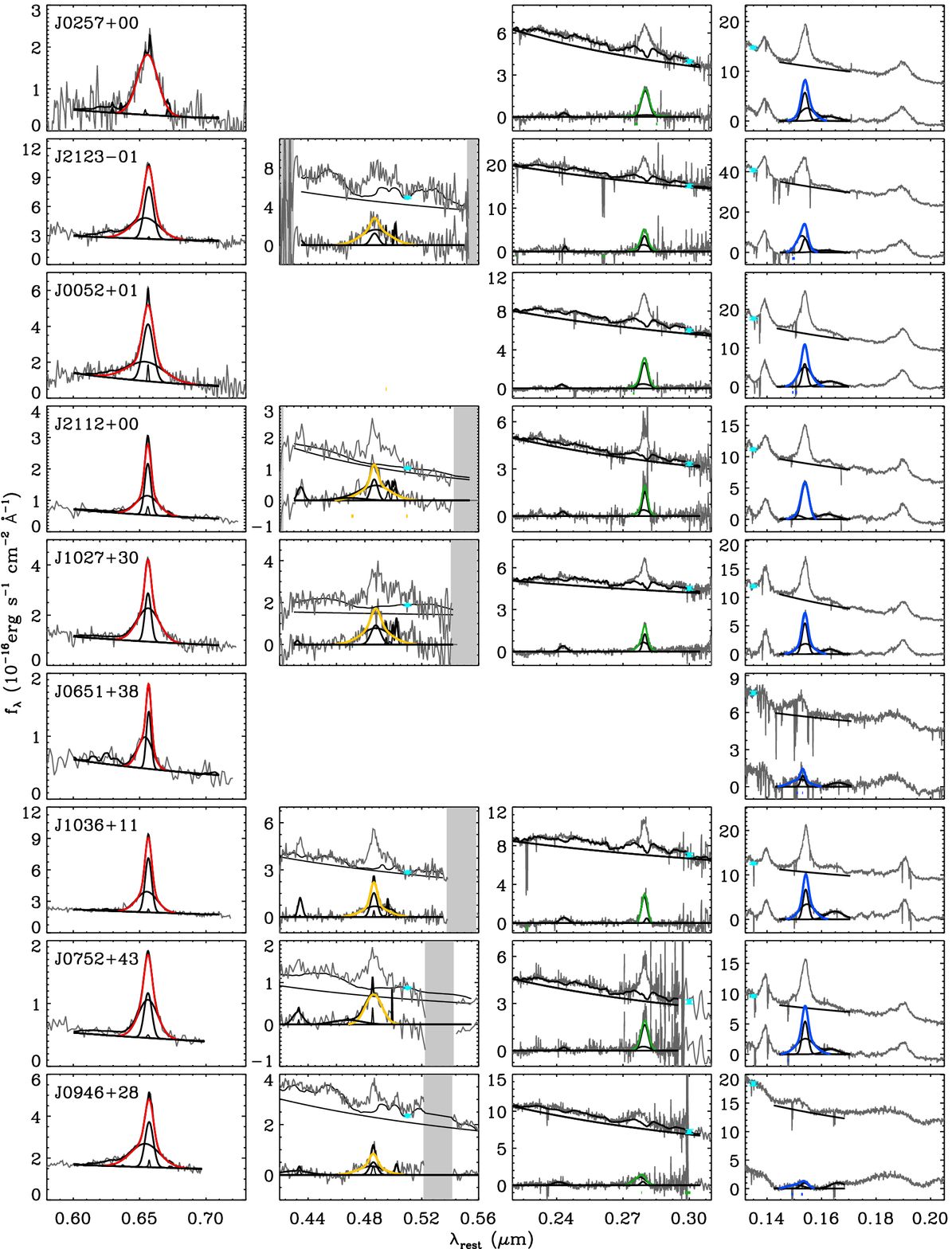}
\end{figure*}
\renewcommand{\tabcolsep}{3pt}
\begin{deluxetable*}{ *{10}{c} }
\tablecolumns{10}
\tabletypesize{\scriptsize}
\tablecaption{Emission line properties}
\tablehead{
\colhead{Name} & \colhead{$v_{\rm H\alpha}-v_{\rm C_{\,IV}}$} & \colhead{FWHM$_{\rm C_{\,III]/IV}}$} & \colhead{FWHM$_{\rm Mg_{\,II}}$} & \colhead{FWHM$_{\rm H\beta}$}  & \colhead{FWHM$_{\rm H\alpha}$} & \colhead{$\sigma_{\rm C_{\,III]/IV}}$} & \colhead{$\sigma_{\rm Mg_{\,II}}$} & \colhead{$\sigma_{\rm H\beta}$}  & \colhead{$\sigma_{\rm H\alpha}$}
}
\startdata
J0102+00 & --- & --- &  3726 $\pm$    96 &  7252 $\pm$   232 &  5686 $\pm$   273 & --- &  2104 $\pm$    45 &  3120 $\pm$    85 &  3186 $\pm$   226\\
J1010+05 & --- & --- &  7653 $\pm$   189 &  8409 $\pm$   311 & $<$ 7613 & --- &  3622 $\pm$    81 &  4938 $\pm$   231 & $<$ 4152\\
J0748+22 & --- & --- &  3581 $\pm$   357 &  7528 $\pm$  1091 &  3401 $\pm$   132 & --- &  1937 $\pm$   142 &  4612 $\pm$  1032 &  2890 $\pm$    61\\
J0840+23 & --- & --- &  6963 $\pm$   181 &  7446 $\pm$  1067 &  7918 $\pm$    72 & --- &  4560 $\pm$   144 &  3162 $\pm$   437 &  3610 $\pm$    21\\
J1057+31 & --- & --- & 13063 $\pm$  3082 & --- & $<$10464 & --- &  6159 $\pm$  1567 & --- & $<$ 5797\\
J0203+13 & --- & --- & 14053 $\pm$  6015 & --- & $<$12747 & --- &  6181 $\pm$  2662 & --- & $<$ 4983\\
J0319-07 & --- & --- &  6687 $\pm$   119 &  5176 $\pm$  1033 &  7431 $\pm$   297 & --- &  4598 $\pm$    97 &  4628 $\pm$   907 &  4522 $\pm$   140\\
J1053+34 & --- & --- &  7910 $\pm$   136 & 14037 $\pm$   633 & $<$11628 & --- &  4263 $\pm$    65 &  5960 $\pm$   271 & $<$ 5129\\
J1035+45 & --- & --- &  4372 $\pm$    56 &  4496 $\pm$   166 &  4279 $\pm$    44 & --- &  3205 $\pm$    15 &  1909 $\pm$    61 &  3697 $\pm$    74\\
J1055+28 &  652 $\pm$  74 &  6266 $\pm$   220 &  5895 $\pm$   241 & 11164 $\pm$  5447 & $<$ 6381 &  5703 $\pm$   234 &  4237 $\pm$   453 &  6156 $\pm$  3235 & $<$ 3730\\
J0146-10 & --- & --- &  6831 $\pm$   225 &  8139 $\pm$   881 &  7789 $\pm$   338 & --- &  5389 $\pm$   128 &  3456 $\pm$   361 &  4205 $\pm$   234\\
J0400-07 & 2007 $\pm$  81 &  6818 $\pm$    89 &  4747 $\pm$   153 &  5551 $\pm$   312 &  4851 $\pm$   203 &  4413 $\pm$    71 &  2719 $\pm$   141 &  5214 $\pm$   220 &  3016 $\pm$   154\\
J0855+05 & 1430 $\pm$  76 &  7832 $\pm$   354 &  7914 $\pm$   291 &  8051 $\pm$  1060 &  7549 $\pm$   245 &  4217 $\pm$   167 &  4322 $\pm$   299 &  4469 $\pm$  1088 &  4164 $\pm$   137\\
J0741+32 &  642 $\pm$ 109 &  6175 $\pm$   134 &  7139 $\pm$   118 &  7586 $\pm$  1008 & $<$ 8281 &  4210 $\pm$    75 &  3823 $\pm$   123 &  4524 $\pm$   857 & $<$ 3598\\
J1522+52 & 6028 $\pm$ 131 & 10546 $\pm$   819 &  6862 $\pm$   436 &  7099 $\pm$   982 &  6514 $\pm$   144 &  5948 $\pm$   553 &  2859 $\pm$   126 &  3650 $\pm$  1503 &  4317 $\pm$    80\\
J1339+11 & 1255 $\pm$  46 &  6214 $\pm$   205 &  4706 $\pm$   209 &  6365 $\pm$  2192 &  5871 $\pm$   222 &  4187 $\pm$   242 &  3626 $\pm$   230 &  3689 $\pm$  1854 &  3704 $\pm$   231\\
J0905+24 &  677 $\pm$  16 &  6183 $\pm$   100 &  3164 $\pm$    97 &  5529 $\pm$   256 &  4782 $\pm$    41 &  3506 $\pm$    85 &  1393 $\pm$    54 &  2659 $\pm$    96 &  3456 $\pm$    42\\
J0257+00 & 1543 $\pm$ 171 &  6813 $\pm$   147 &  4771 $\pm$   113 & --- & $<$ 8191 &  4181 $\pm$   178 &  3103 $\pm$   212 & --- & $<$ 3142\\
J2123-01 & 2244 $\pm$  51 &  7282 $\pm$   125 &  4123 $\pm$   230 &  8006 $\pm$  1362 &  4649 $\pm$   119 &  3745 $\pm$    62 &  2476 $\pm$   247 &  5844 $\pm$  1262 &  3510 $\pm$   106\\
J0052+01 & 1961 $\pm$  44 &  6327 $\pm$    93 &  4255 $\pm$   157 & --- &  5179 $\pm$   174 &  4258 $\pm$    74 &  2511 $\pm$   324 & --- &  4048 $\pm$   105\\
J2112+00 & 1859 $\pm$  28 &  6724 $\pm$   103 &  3232 $\pm$   181 &  5336 $\pm$   725 &  3117 $\pm$   113 &  3394 $\pm$   106 &  2006 $\pm$   258 &  5334 $\pm$   702 &  2814 $\pm$   110\\
J1027+30 & 1846 $\pm$  39 &  5868 $\pm$   159 &  3265 $\pm$   193 &  7394 $\pm$   900 &  3686 $\pm$   139 &  4212 $\pm$   289 &  2177 $\pm$   160 &  5508 $\pm$   854 &  3029 $\pm$    87\\
J0651+38 & 3651 $\pm$  90 &  5393 $\pm$   275 & --- & --- &  2593 $\pm$   180 &  5394 $\pm$   270 & --- & --- &  1952 $\pm$    87\\
J1036+11 & 1236 $\pm$  23 &  5915 $\pm$   127 &  3412 $\pm$    90 &  4433 $\pm$   422 &  3299 $\pm$    55 &  4342 $\pm$   155 &  1399 $\pm$    30 &  4250 $\pm$   471 &  2632 $\pm$    58\\
J0752+43 & 1700 $\pm$  47 &  6379 $\pm$    68 &  4211 $\pm$   178 &  8517 $\pm$   684 &  4379 $\pm$   338 &  5052 $\pm$    74 &  2040 $\pm$   257 &  3617 $\pm$   269 &  2722 $\pm$   152\\
J0946+28 & 5320 $\pm$ 225 & 11172 $\pm$  1870 &  5901 $\pm$   536 &  5666 $\pm$   735 &  4459 $\pm$    92 &  5584 $\pm$   635 &  2769 $\pm$   189 &  4738 $\pm$   453 &  3881 $\pm$    74
\enddata
\tablecomments{The \ion{C}{4} to H$\alpha$ broad line shift (positive for blueshifted \ion{C}{4}, \kms) and the line widths (FWHM and $\sigma$, in \kms). The \ion{C}{3}] line widths are used instead of the \ion{C}{4} when the \ion{C}{4} line is not covered. Upper limits to the line width values are associated with double-peaked H$\alpha$ emission.}
\end{deluxetable*}

\renewcommand{\tabcolsep}{2.5pt}
\begin{deluxetable*}{ *{10}{c} }
\tablecolumns{10}
\tabletypesize{\scriptsize}
\tablecaption{Luminosities and $M_{\rm BH}$ values}
\tablehead{
\colhead{Name} & \colhead{log $L_{1350/1450}$} & \colhead{log $L_{3000}$} & \colhead{log $L_{5100}$} & \colhead{log $L_{\rm H\alpha}$} & \colhead{$M_{\rm BH,C_{\,III]/IV}}$} & \colhead{$M_{\rm BH,Mg_{\,II}}$} & \colhead{$M_{\rm BH,H\beta}$} & \colhead{$M_{\rm BH,H\alpha}$} & \colhead{Flags}
}
\startdata
J0102+00 & --- & 45.86 $\pm$ 0.009 & 45.68 $\pm$ 0.008 & 44.34 $\pm$ 0.042 & --- & 9.03 $\pm$ 0.14 & 9.56 $\pm$ 0.13 & 9.53 $\pm$ 0.14 & \\
J1010+05 & --- & 46.16 $\pm$ 0.010 & 45.95 $\pm$ 0.008 & 44.67 $\pm$ 0.052 & --- & 9.95 $\pm$ 0.16 & 9.83 $\pm$ 0.14 & $<$9.95 &  DPE\\
J0748+22 & --- & 46.62 $\pm$ 0.009 & 46.32 $\pm$ 0.009 & 44.83 $\pm$ 0.025 & --- & 9.40 $\pm$ 0.19 & 9.93 $\pm$ 0.19 & 9.40 $\pm$ 0.15 & \\
J0840+23 & 46.91 $\pm$ 0.029 & 46.67 $\pm$ 0.010 & 46.33 $\pm$ 0.008 & 45.06 $\pm$ 0.010 & 10.07 $\pm$ 0.21 & 10.13 $\pm$ 0.16 & 9.93 $\pm$ 0.19 & 10.19 $\pm$ 0.14 & \\
J1057+31 & 46.11 $\pm$ 0.068 & 46.06 $\pm$ 0.015 & 45.89 $\pm$ 0.058 & 44.83 $\pm$ 0.014 & 9.06 $\pm$ 0.22 & 10.46 $\pm$ 0.29 & --- & $<$10.21 &  DPE, \ion{Fe}{2}\\
J0203+13 & 45.92 $\pm$ 0.045 & 45.97 $\pm$ 0.012 & 45.81 $\pm$ 0.013 & 44.84 $\pm$ 0.023 & 9.70 $\pm$ 0.21 & 10.49 $\pm$ 0.48 & --- & $<$10.35 &  DPE, \ion{Fe}{2}\\
J0319-07 & 46.60 $\pm$ 0.029 & 46.44 $\pm$ 0.010 & 46.11 $\pm$ 0.009 & 44.87 $\pm$ 0.033 & 9.85 $\pm$ 0.20 & 9.96 $\pm$ 0.16 & 9.52 $\pm$ 0.22 & 10.01 $\pm$ 0.14 & [\ion{O}{3}]\\
J1053+34 & 46.62 $\pm$ 0.027 & 46.51 $\pm$ 0.008 & 46.20 $\pm$ 0.008 & 45.11 $\pm$ 0.054 & 10.08 $\pm$ 0.22 & 10.17 $\pm$ 0.16 & 10.41 $\pm$ 0.14 & $<$10.47 &  DPE, \ion{Fe}{2}\\
J1035+45 & 46.70 $\pm$ 0.049 & 46.84 $\pm$ 0.010 & 46.68 $\pm$ 0.008 & 45.46 $\pm$ 0.012 & 10.08 $\pm$ 0.21 & 9.73 $\pm$ 0.16 & 9.67 $\pm$ 0.15 & 9.81 $\pm$ 0.15 & [\ion{O}{3}]\\
J1055+28 & 46.90 $\pm$ 0.076 & 46.54 $\pm$ 0.012 & 46.22 $\pm$ 0.010 & 44.73 $\pm$ 0.055 & 9.94 $\pm$ 0.21 & 9.88 $\pm$ 0.16 & 10.27 $\pm$ 0.45 & $<$9.93 &  DPE, [\ion{O}{3}]\\
J0146-10 & 46.12 $\pm$ 0.048 & 46.31 $\pm$ 0.011 & 46.17 $\pm$ 0.008 & 44.88 $\pm$ 0.051 & 9.34 $\pm$ 0.19 & 9.91 $\pm$ 0.16 & 9.92 $\pm$ 0.17 & 10.09 $\pm$ 0.15 & \ion{Fe}{2}, [\ion{O}{3}]\\
J0400-07 & 46.86 $\pm$ 0.038 & 46.63 $\pm$ 0.009 & 46.33 $\pm$ 0.008 & 44.96 $\pm$ 0.041 & 10.00 $\pm$ 0.21 & 9.70 $\pm$ 0.16 & 9.68 $\pm$ 0.15 & 9.73 $\pm$ 0.15 & \ion{C}{4}\\
J0855+05 & 46.70 $\pm$ 0.128 & 46.46 $\pm$ 0.013 & 46.11 $\pm$ 0.010 & 44.89 $\pm$ 0.031 & 10.04 $\pm$ 0.22 & 10.15 $\pm$ 0.16 & 9.87 $\pm$ 0.18 & 10.02 $\pm$ 0.14 & \\
J0741+32 & 46.50 $\pm$ 0.008 & 46.33 $\pm$ 0.010 & 45.94 $\pm$ 0.014 & 44.67 $\pm$ 0.262 & 9.71 $\pm$ 0.20 & 9.97 $\pm$ 0.16 & 9.73 $\pm$ 0.18 & $<$10.02 &  DPE\\
J1522+52 & 47.61 $\pm$ 0.008 & 47.51 $\pm$ 0.013 & 47.19 $\pm$ 0.008 & 45.74 $\pm$ 0.019 & 10.81 $\pm$ 0.24 & 10.57 $\pm$ 0.19 & 10.34 $\pm$ 0.20 & 10.47 $\pm$ 0.16 & \ion{C}{4}\\
J1339+11 & 47.02 $\pm$ 0.008 & 46.83 $\pm$ 0.023 & 46.50 $\pm$ 0.009 & 45.12 $\pm$ 0.040 & 10.00 $\pm$ 0.21 & 9.80 $\pm$ 0.17 & 9.88 $\pm$ 0.33 & 10.00 $\pm$ 0.15 & \\
J0905+24 & 46.95 $\pm$ 0.008 & 46.93 $\pm$ 0.011 & 46.74 $\pm$ 0.008 & 45.49 $\pm$ 0.011 & 9.96 $\pm$ 0.21 & 9.44 $\pm$ 0.17 & 9.88 $\pm$ 0.15 & 9.94 $\pm$ 0.15 & [\ion{O}{3}]\\
J0257+00 & 46.89 $\pm$ 0.008 & 46.67 $\pm$ 0.013 & 46.22 $\pm$ 0.029 & 45.03 $\pm$ 0.028 & 10.01 $\pm$ 0.21 & 9.73 $\pm$ 0.16 & --- & $<$10.16 &  DPE\\
J2123-01 & 47.34 $\pm$ 0.008 & 47.26 $\pm$ 0.010 & 47.00 $\pm$ 0.009 & 45.62 $\pm$ 0.023 & 10.32 $\pm$ 0.22 & 9.89 $\pm$ 0.18 & 10.37 $\pm$ 0.21 & 10.05 $\pm$ 0.16 & \ion{C}{4}\\
J0052+01 & 46.98 $\pm$ 0.008 & 46.86 $\pm$ 0.011 & 46.55 $\pm$ 0.024 & 45.44 $\pm$ 0.024 & 9.99 $\pm$ 0.21 & 9.71 $\pm$ 0.17 & --- & 9.91 $\pm$ 0.15 & \\
J2112+00 & 46.81 $\pm$ 0.008 & 46.62 $\pm$ 0.013 & 46.35 $\pm$ 0.010 & 44.98 $\pm$ 0.028 & 9.96 $\pm$ 0.20 & 9.29 $\pm$ 0.17 & 9.66 $\pm$ 0.18 & 9.34 $\pm$ 0.15 & \\
J1027+30 & 46.84 $\pm$ 0.008 & 46.76 $\pm$ 0.013 & 46.61 $\pm$ 0.010 & 45.23 $\pm$ 0.028 & 9.85 $\pm$ 0.20 & 9.38 $\pm$ 0.17 & 10.11 $\pm$ 0.18 & 9.63 $\pm$ 0.15 & \\
J0651+38 & 46.65 $\pm$ 0.008 & --- & 46.31 $\pm$ 0.069 & 44.68 $\pm$ 0.040 & 9.66 $\pm$ 0.20 & --- & --- & 9.15 $\pm$ 0.15 & \ion{C}{4}\\
J1036+11 & 46.87 $\pm$ 0.008 & 46.97 $\pm$ 0.012 & 46.80 $\pm$ 0.008 & 45.49 $\pm$ 0.016 & 9.87 $\pm$ 0.20 & 9.53 $\pm$ 0.17 & 9.72 $\pm$ 0.17 & 9.63 $\pm$ 0.15 & \\
J0752+43 & 46.80 $\pm$ 0.008 & 46.65 $\pm$ 0.018 & 46.33 $\pm$ 0.012 & 44.90 $\pm$ 0.063 & 9.90 $\pm$ 0.20 & 9.59 $\pm$ 0.16 & 10.03 $\pm$ 0.16 & 9.64 $\pm$ 0.16 & \\
J0946+28 & 47.08 $\pm$ 0.008 & 47.02 $\pm$ 0.017 & 46.76 $\pm$ 0.008 & 45.36 $\pm$ 0.016 & 10.57 $\pm$ 0.27 & 10.14 $\pm$ 0.19 & 9.94 $\pm$ 0.19 & 9.89 $\pm$ 0.15 & \ion{C}{4}
\enddata
\tablecomments{The monochromatic continuum and broad line luminosities (\ergs), and $M_{\rm BH}$ values ($M_{\odot}$). Upper limits to the $M_{\rm BH}$ values are associated with double-peaked H$\alpha$ emission marked as DPE in the Flags column. Extremely broad ($>20,000\kms$) \ion{Fe}{2}, highly blueshifted ($>2000\kms$) \ion{C}{4}, and broad ($>1000\kms$) [\ion{O}{3}], are flagged \ion{Fe}{2}, \ion{C}{4}, and [\ion{O}{3}], respectively.
}
\end{deluxetable*}

Lastly, we fit the rest-frame 1445--1705\,\AA\ region around the \ion{C}{4} emission. When the \ion{C}{4} was unavailable or severely absorbed, we fit the rest-frame 1670--2050\,\AA\ region around \ion{C}{3}] to use the FWHM as a surrogate to that of \ion{C}{4} (S12). We fixed the height and width of the \ion{Fe}{2} complex from those nearby the \ion{Mg}{2} because this feature is weaker around the \ion{C}{4} and \ion{C}{3}]. We used a power-law continuum and double broad Gaussians for the \ion{C}{4} and \ion{C}{3}]. For the \ion{C}{4} region we used single broad Gaussians to fit the 1600\,\AA\ feature \citep{Lao94}, and the blended \ion{He}{2} and \ion{O}{3}] emission around 1650\,\AA. For the \ion{C}{3}] region we used single broad Gaussians to model each of \ion{N}{4}$\lambda$1718, \ion{N}{3}]$\lambda$1750, \ion{Fe}{2}$\lambda$1786 (UV191), \ion{Si}{2}$\lambda$1816, \ion{Al}{3}$\lambda$1857, and \ion{Si}{3}]$\lambda$1891. We masked or changed the fitting range of the \ion{C}{4}/\ion{C}{3}] spectra showing strong absorption features, and clipped the spectra showing weaker absorption features by redoing the fit after removing the data below 2.5-$\sigma$ from the fit. We find the \ion{C}{3}] centers to lie between $-$2000 and 1000\,\kms\ of the H$\alpha$ redshift, but \ion{C}{4} is often more blueshifted, so its centers were set between $-$8000 and 2000\,\kms. Depending on the spectral coverage, either $L_{1350}$ or $L_{1450}$ was calculated based on their similarity \citep{Ves06} as is expected from the small lever arm between them. When the 1350\,\AA\ was available we computed the error weighted average of the 1340--1360\,\AA\ fluxes, and if only 1450\,\AA\ was covered, the 1440--1460\,\AA\ fluxes were used as an approximate measure of $L_{1350}$. When neither were available, we extrapolated the continuum and \ion{Fe}{2} emission from the \ion{C}{3}] fit down to 1450\,\AA\ while propagating the errors from the fit.

In Figure 3 we plot the fit to the spectra around the broad emission lines for each object, and in Tables 2--3 we summarize the spectral measurements. Once the spectral fitting was performed around each broad line, we derived the broad line FWHM and line dispersion (hereafter $\sigma$). The errors on FWHM were determined by equating the FWHM of the combined double Gaussian model as a linear combination of the constituent single Gaussian FWHMs (interpolated or extrapolated, depending on the relative magnitude of the FWHMs), and propagating the errors from each single FWHM measurement. For $\sigma$ and its error, we used the second moment of the model fit fluxes up to the point they are equal to the flux errors; this corresponds to $\pm 2.0$\,FWHM from the broad line center on average. Out of the 26 objects in our sample, we compile 21, 26, 25, and 23 line widths from H$\beta$, H$\alpha$, \ion{Mg}{2}, and \ion{C}{4}/\ion{C}{3}], respectively. We note that some of the broad component's FWHM values are close to the lower limit of 2000\,\kms\ and are intermediate in width, e.g., H$\alpha$ in J1057+31 and \ion{Mg}{2} in J1053+34. These could be confused with narrow lines with strong outflows, where we test the possible change in the $M_{\rm BH}$ values in section 4.2.4. Meanwhile, the spectroscopic continuum luminosities are corrected for the photometric calibration uncertainty ($\sim$\,0.02 mag) involved when scaling the spectra (section 2.2).

\section{Results}
\subsection{$M_{\rm BH}$ estimation of our sample}
The determined continuum luminosities and broad line FWHMs are plugged into single-epoch $M_{\rm BH}$ estimators for AGNs from J15 for each emission line measurement, assuming the constant $f=5.1\pm1.3$ from \citet{Woo13} and the $R_{\rm BLR }$--$L$ relation from \citet{Ben13}. 
\begin{eqnarray}\begin{aligned}
\log \Big(&\frac{M_{\rm BH}}{M_{\odot}}\Big)=a+\log\Big(\frac{f}{5.1}\Big)\\
&+b\,\log\Big(\frac{L}{\rm 10^{44} \ergs}\Big)+c\,\log\Big(\frac{\mathrm{FWHM}}{\rm10^{3}\kms}\Big),
\end{aligned}\end{eqnarray}
where $(a, b, c)$ values using the combination of ($L$, FWHM) below are
\begin{eqnarray}\hspace{-10pt}\begin{aligned}
&(a, b, c)=\\
&(L_{5100}, \rm{FWHM}_{\rm H\beta}): (6.94\pm0.12, 0.533\pm0.034, 2)\\
&(L_{5100}, \rm{FWHM}_{\rm H\alpha}): (7.05\pm0.12, 0.533\pm0.034, 2.12\pm0.03)\\ 
&(100L_{\rm H\alpha}, \rm{FWHM}_{\rm H\alpha}): (6.72\pm0.12, 0.511\pm0.033, 2.12\pm0.03)\\ 
&(L_{3000}, \rm{FWHM}_{\rm Mg_{\,II}}): (6.62\pm0.12, 0.548\pm0.035, 2.45\pm0.06)\\ 
&(L_{1350}, \rm{FWHM}_{\rm C_{\,IV}}): (6.67\pm0.15, 0.547\pm0.037, 2.11\pm0.11).
\end{aligned}\end{eqnarray}
Because some of the Balmer lines covered by the IRTF spectra have marginal sensitivity and the H$\beta$ line is a few times weaker than H$\alpha$, we compare the intrinsic scatter $\sigma_{\rm int}$\footnote{$\sigma_{\rm int}^{2}=\Sigma_{i=1}^{N}\{(y_{i}-x_{i})^{2}-(\Delta x_{i}^{2}+\Delta y_{i}^{2})\}/(N-1)$ for $N$ measurements ($x_{i},y_{i}$) and errors ($\Delta x_{i}, \Delta y_{i}$).} between the H$\alpha$ and H$\beta$ line based $M_{\rm BH}$ values, for the continuum sensitivity bins $5<\rm S/N_{\rm H\beta}<10$ and $\rm S/N_{\rm H\beta}$\footnote{The continuum S/N values for the denoted line in subscript letters are the median calculated over the wavelengths used to fit the line region.}$>10$. We find $\sigma_{\rm int}=0.21$\,dex and $\sigma_{\rm int}<0$ for the two sensitivity bins so that larger systematic uncertainties are affecting the $M_{\rm BH}$ values at lower sensitivity (e.g., \citealt{Den09}). This effect should be negligible in the \ion{Mg}{2}, \ion{C}{4}, or \ion{C}{3}] masses derived under much higher sensitivity, all at $\rm S/N>35$. We thus limit the usage of H$\beta$ $M_{\rm BH}$ values to $\rm S/N_{\rm H\beta}>10$, and average with the H$\alpha$-based masses when used for comparison with the rest-UV $M_{\rm BH}$ values.

\begin{figure*}
\centering
\includegraphics[scale=.665]{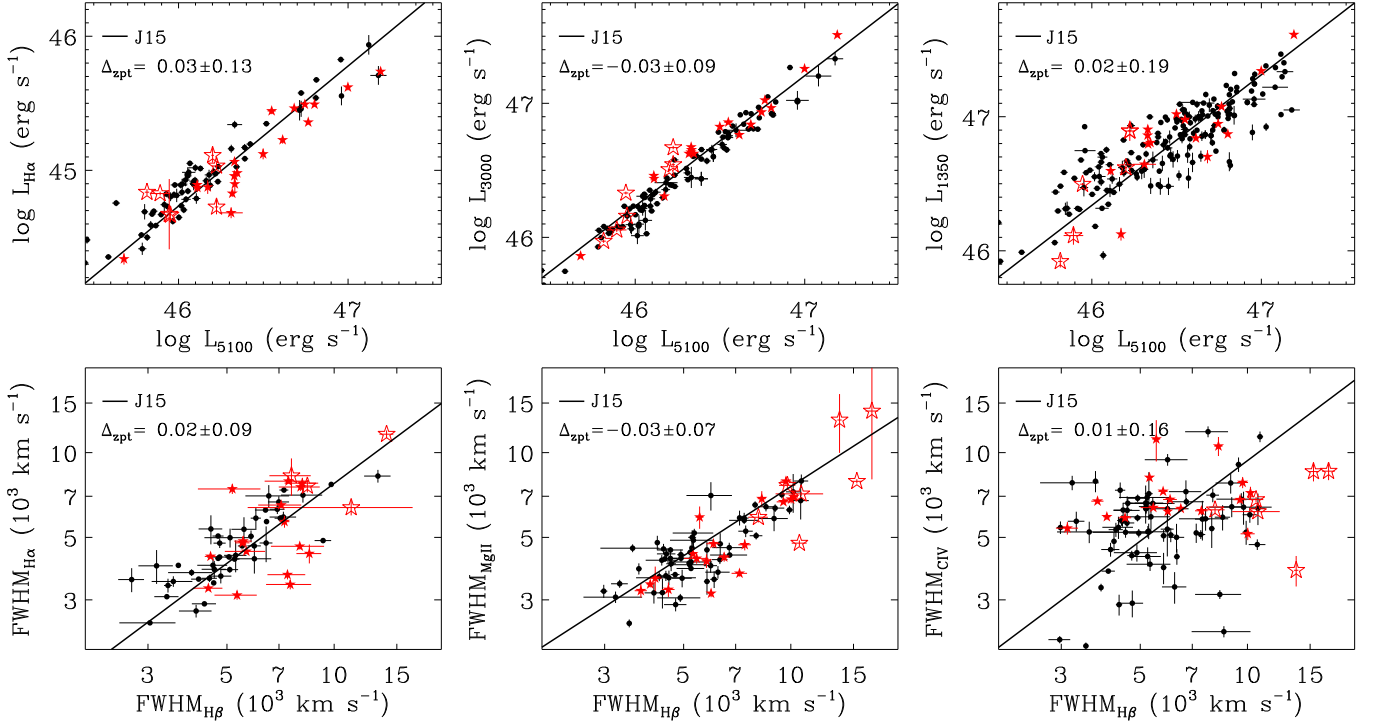}
\caption{The continuum--line luminosity relations (top) and line width relations (bottom), for type-1 AGNs at the luminous and massive end. Among the IRTF data (red stars), those double-peaked in H$\alpha$ (red open stars) are highlighted. The luminosities and line widths plotted are limited to having within 20\% uncertainty, except for the double-peaked emitters. We compile the data from references spanning similarly luminous quasars as our sample (black dots, S12; J15; \citealt{She16}), removing $\rm FWHM_{C_{\,IV}}$ value affected by broad absorption. The zeropoint offset of the combined data (excluding the double-peaked emitters) with respect to the J15 relation (solid lines) and intrinsic scatter are denoted as $\Delta_{\text{zpt}}$. The FWHM$_{\rm H\beta}$ for the bottom right two panels are converted from FWHM$_{\rm H\alpha}$ whenever available using the J15 relation, in order to benefit from the enhanced sensitivity of the H$\alpha$ line.} 
\end{figure*}

\begin{figure*}
\centering
\includegraphics[scale=.665]{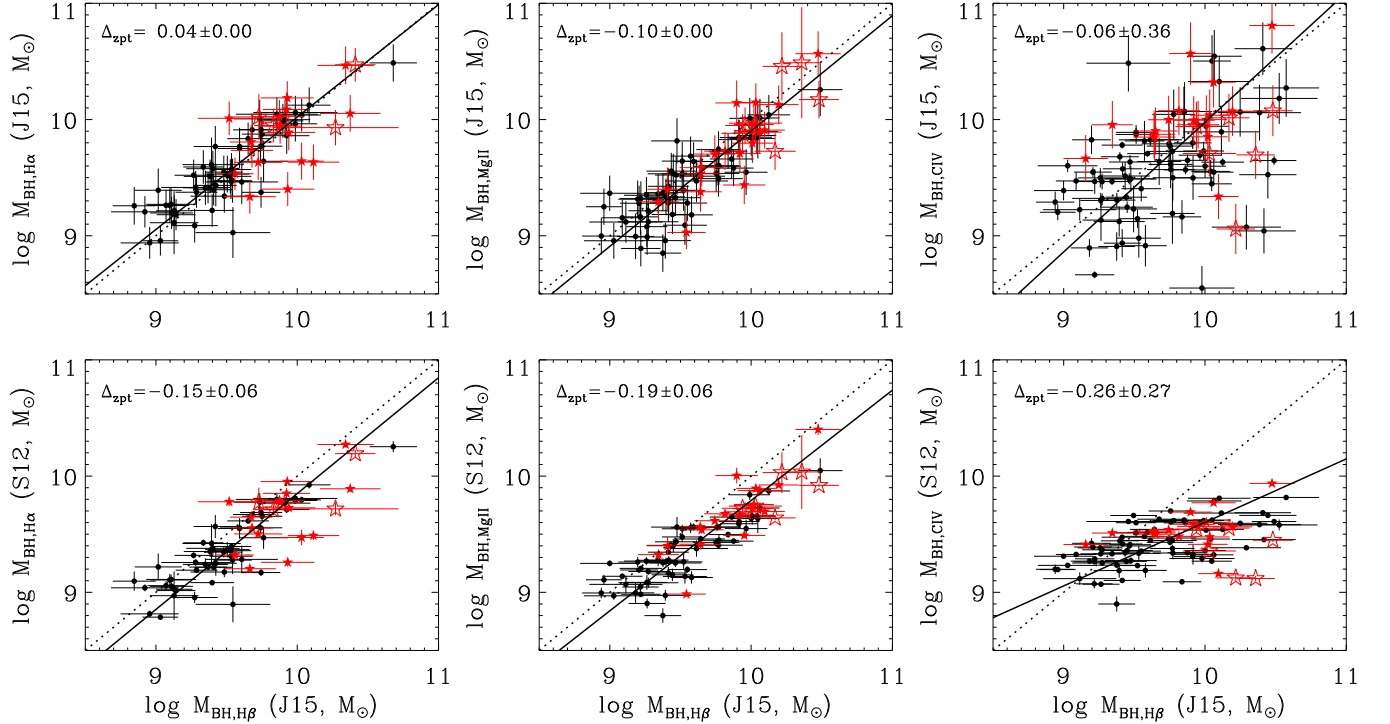}
\caption{The massive end $M_{\rm BH}$ relations out of various lines using the J15 relation (top) and comparison of $M_{\rm BH}$ estimates between J15 and S12 (bottom), with S12 estimators corrected to have the same constant factor as the J15 ($f$=5.1). The data, colors, and the symbols follow those of Figure 4, while the $M_{\rm BH}$ values plotted are limited to having within 0.3\,dex uncertainty except for the double-peaked emitters. The zeropoint offset of the combined data with respect to a one-to-one relation and intrinsic scatter (fixed to zero when negative) are denoted as $\Delta_{\text{zpt}}$. The linear fit to the data and a one-to-one relation are shown in solid and dotted lines respectively. The double-peaked emitters are excluded for the $\Delta_{\text{zpt}}$ calculation and the linear fit. The $M_{\rm BH, H\beta}$ for the center and right panels are converted from FWHM$_{\rm H\alpha}$ whenever available using the J15 relation, in order to benefit from the enhanced sensitivity of the H$\alpha$ line.} 
\end{figure*}

The estimators from J15 were calibrated to yield consistent rest-UV to rest-optical $M_{\rm BH}$ values over a wide range of luminosity and redshift, suitable for this study. We further examine where the measured continuum luminosities and broad line FWHMs of our sample with extremely large masses, fall with respect to the quasars with similar luminosities. In Figures 4--5, we plot the continuum--line luminosity relations, FWHM relations, and the $M_{\rm BH}$ relations based on the H$\alpha$, H$\beta$, \ion{Mg}{2}, and \ion{C}{4} lines, from this work and existing observations of similarly luminous quasars (S12, J15, \citealt{She16}). The offset and $\sigma_{\rm int}$ of the combined data with respect to the best-fit relations in previous works are printed on each panel. From Figure 4 we find that our objects together with similarly luminous quasars, follow the luminosity and line width relations of J15. The data shows negligible offset, and $\sigma_{\rm int}$ similar to that of the J15 relation, demonstrating that the J15 calibrations are useful even for extremely massive AGNs. We note that the extremely massive AGNs are mostly from this work and they distribute similar in luminosity space to other luminous quasars, but have FWHM values higher than other luminous quasars. This suggests that the main factor that give rise to EMBH estimates is their wide velocity widths. 

Also, we check in Figure 5 whether the rest-UV to rest-optical $M_{\rm BH}$ values are mutually consistent at the massive end, using the J15 and S12 relations. The H$\beta$ and H$\alpha$ $M_{\rm BH}$ values of luminous, massive AGNs are consistent with each other irrespective of using the S12 or J15 estimators, albeit with a smaller intrinsic scatter for the J15 estimator due to the inclusion of the measurement uncertainties in the $f$-factor, $R_{\rm BLR }$--$L$ relation, and luminosity/line width correlations. The rest-UV to Balmer $M_{\rm BH}$ values for EMBHs are mutually consistent using the J15 estimators, whereas the S12 estimators lead to systematically underestimated S12 rest-UV to J15 Balmer $M_{\rm BH}$ ratios ($\sim$ 0.21 and 0.40 dex underestimation in \ion{Mg}{2} and \ion{C}{4} $M_{\rm BH}$ values respectively at $M_{\rm BH, H\beta}=10^{10}M_{\odot}$, and more deviations at higher $M_{\rm BH}$). The existing estimators determined from a relatively limited dynamic range in rest-UV FWHM tend to have shallower scaling of the FWHM into the mass estimator compared to the J15, underestimating the rest-UV $M_{\rm BH}$ values at the massive end. Therefore, we keep the J15 estimator as a relatively more reliable $M_{\rm BH}$ indicator for the rest of the paper.

\subsection{Mass biasing factors in the spectra}
\subsubsection{Double-peaked broad emission}
We find seven out of 26 broad H$\alpha$ profiles classified as double-peaked emitters in section 3 (J0203+13, J0257+00, J0741+32, J1010+05, J1053+34, J1055+28, J1057+31), best fit as triple Gaussians with clear blue- and redshifted components comparable in strength or dominating over the central broad component. In Figure 6, we plot the H$\alpha$ fit and the noise spectrum of the double-peaked emitters, and find that the double-peaked features are stronger than the noise levels. We follow \citet{Bas05} to place these objects along various parameters describing its profile: the shape, $\rm (FW1/4M+FW3/4M)/(2FWHM)$; the asymmetry, $\rm (\lambda_{3/4}-\lambda_{1/4})/FWHM$; and the shift, $\rm (\lambda_{3/4}-\lambda_{4/4})/FWHM$, where FWN/4M and $\lambda_{\rm N/4}$ (N=1--4) are the width and centroid of the Nth-quarter maximum of the line. The shape parameter for the seven double-peaked objects ranges within $0.89$--$1.16$ (1.03 on average), shifted to the distribution of $1.02$--$1.37$ (1.16 on average) for the non-double-peaked objects and lying on the smaller end of the H$\beta$ distribution from \citet{Bas05}. The distribution of asymmetry and shift parameters are indistinguishable between the double-peaked and ordinary profiles. Overall, our selected double-peaked profiles are systematically different to ordinary profiles as being wider towards the peak or the wings. However, we still make cautions for directly comparing our double-peaked profiles to those in the literature, as some examples (J1010+05, J1055+28) look marginal in appearance.

\begin{figure*}
\centering
\includegraphics[scale=.95]{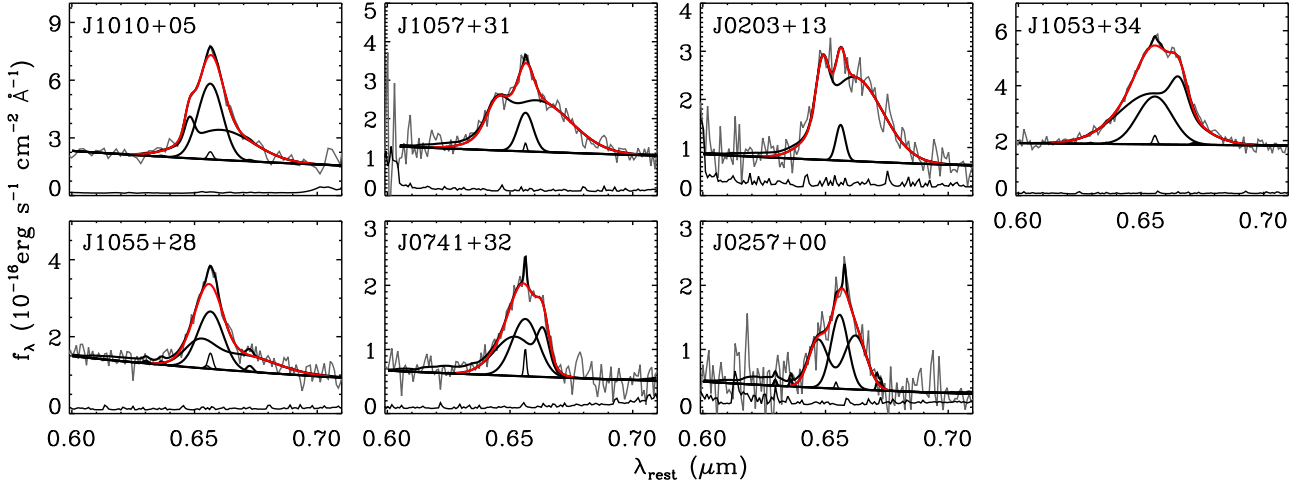}
\caption{The H$\alpha$ profiles of double-peaked emitters identified using triple broad Gaussian fitting. The combined blue- and redshifted double-peaked broad emission and the central single broad emission components are separately plotted (thick black), as well as the total broad (red) and the narrow emission (thin black). The noise spectrum is shown below the object spectrum.} 
\end{figure*}

From Figure 4, we find the double-peaked emitters are slightly above the J15 $L_{5100}$--$L_{\rm{H}\alpha}$ relation by 0.068\,dex on average, though within $\sigma_{\rm int}=0.095$\,dex from the J15 relation. Also, the double-peaked emitters lie below the $\rm FWHM_{H\beta}$--$\rm FWHM_{C_{\,IV}}$ relation by $-0.24$\,dex on average, slightly larger than $\sigma_{\rm int}=0.21$\,dex from the J15 relation. We further estimate the differences in the H$\alpha$ and \ion{C}{4} $M_{\rm BH}$ estimates for the double-peaked emitters, to find $\log M_{\rm BH, H\alpha}/M_{\rm BH, C_{\,IV}}$=$-0.01$--$1.15$ ($0.45$ on average). The \ion{C}{4} spectra of double-peaked H$\alpha$ emitters appear to show weaker double-peaks (e.g., \citealt{Era04}), which is consistent with the larger $\rm FWHM_{H\alpha}$ to $\rm FWHM_{C_{\,IV}}$ ratios for our double-peaked emitters, $0.01$--$0.44$ ($0.16$ on average) dex.

We independently check if using the line widths for double-peaked H$\alpha$ emitters lead to overestimated $M_{\rm BH}$ values (e.g,. \citealt{Wu04}; \citealt{Zha07}), from the stellar velocity dispersion ($\sigma_{*}$) and $\rm FWHM_{ H\alpha}$ measurements of 10 double-peaked emitters from \citet{Lew06}, where we find 9 objects having both $\sigma_{*}$ and $\rm FWHM_{ H\alpha}$ compiled from the literature (\citealt{Era93}; \citealt{Era94}; \citealt{Bar02}; \citealt{Ser02}; \citealt{Nel04}; \citealt{Lew06}; \citealt{Lew10}). Applying the $M_{\rm BH}$--$\sigma_{*}$ relations from \citet{Kor13} and \citet{McC13} give a range of $M_{\rm BH} (\sigma_{*})$ values, considering that the slope of the relation is different between the references. Meanwhile, we converted the bolometric luminosities in \citet{Lew06} to $L_{5100}$ using the bolometric correction 10.33 from \citet{Ric06}, to estimate the $M_{\rm BH, H\alpha}$ using the J15 estimator. We use only the objects with $L_{5100}>10^{41.5}$\ergs\ where the mass calibration is defined, yielding six $M_{\rm BH}$ estimates. To minimize the time dependent changes in $\rm FWHM_{ H\alpha}$, we used the averaged value through monitored observations available for four objects. Assuming that the double-peaked emitters are lying on the $M_{\rm BH}$--$\sigma_{*}$ relation, we find the $M_{\rm BH, H\alpha}$ values are larger than the $M_{\rm BH} (\sigma_{*})$ values by 0.32--1.14 (0.85 on average) dex, and 0.32--2.06 (1.01 on average) dex, out of the \citet{Kor13} and \citet{McC13} relations respectively. This is in agreement with the \citet{Zha07} result.

We make a cautionary note that the $M_{\rm BH}$ estimates thought to be less affected by the double-peaked features involve large intrinsic scatter ($\sim$\,0.4\,dex for the $M_{\rm BH, C_{\,IV}}$ from J15, $\sim$\,0.3--0.4\,dex for the $M_{\rm BH} (\sigma_{*})$ from \citealt{Kor13}; \citealt{McC13}), and the bolometric luminosities in \citet{Lew06} suffer from incomplete wavelength coverage. Still, the overestimation in $M_{\rm BH}$ comparable to or even more sizable than the uncertainties, suggests that the $M_{\rm BH}$ values of double-peaked emitters using the full broad Balmer emission profile, are possibly overestimated. We find that four objects out of the seven double-peaked H$\alpha$ emitters with H$\beta$ S/N\,$>$\,5 have relatively broad $\rm FWHM_{H\beta}$ values of 7600--14,000 (10,300 on average) \kms, compared to 4430--8500 (6580 on average) \kms\ for the rest of the sample. Our H$\beta$ observations have poorer sensitivities than the H$\alpha$ to identify double-peaked emission, but the FWHMs are consistent with the expected broadening of the line profile from a rotating accretion disk.

\subsubsection{Extremely wide \ion{Fe}{2} solution near \ion{Mg}{2}}
Next, we investigate the effect of \ion{Fe}{2} subtraction on the systematic uncertainty of the $M_{\rm BH}$ estimates. Whereas most of the H$\beta$ spectral fits in section 3 yielded a least-squares solution for the \ion{Fe}{2} complex\footnote{The exceptions are the \ion{Fe}{2} fitted with the narrowest widths (900\,\kms\ from the template, J0748+22, J0905+24), but they are acceptable considering that the \ion{Fe}{2} of these objects are too weak to be well constrained.}, four out of 25 \ion{Mg}{2} spectra (J0146-10, J0203+13, J1053+34, J1057+31) did not converge until the $\rm FWHM_{Fe\,II}$ reached its maximum limit of 20,000\,\kms, with the first three classified as double-peaked emitters from the H$\alpha$ spectra. We note that the automated fitting of SDSS spectra from \citet{She11} also identifies extremely broad \ion{Fe}{2} solutions within our sample (J1010+05, J1053+34, J1057+31, J1522+52), with the first three double-peaked in our H$\alpha$. To improve the \ion{Mg}{2} fit of these sources we attempted to include a Balmer continuum emission component that is usually degenerate with the power-law continuum and the \ion{Fe}{2} complex (e.g., \citealt{Mao93}; \citealt{Wan09}), following the functional form and parameter boundaries of S12. Fitting both the power-law and Balmer continuua does not reduce the extremely wide \ion{Fe}{2} widths to convergence for all four of our sources however, such that either the standard \ion{Fe}{2} template does not fit these quasar spectra, or the broad Gaussian model is insufficient to model these \ion{Mg}{2} profiles. In any case, the $M_{\rm BH}$ measurements associated with extremely broad \ion{Fe}{2} solutions require careful interpretation as they correlate with sources showing double-peaked H$\alpha$ emission.

\subsubsection{Blueshifted \ion{C}{4}}
Third, though it is typical to find \ion{C}{4} emission in quasar spectra blueshifted relative to the rest-optical or \ion{Mg}{2} redshifts by $\sim$1000\,\kms\ (e.g., \citealt{Ric02}; \citealt{She08}; \citealt{She11}), we find stronger blueshifts in our sample. For comparison, the mean and rms scatter of the \ion{C}{4} to \ion{Mg}{2} blueshift from SDSS DR7 quasars at $1.6<z<2.0$ with ${\rm S/N}>20$ in both \ion{Mg}{2} and \ion{C}{4} is $844 \pm 916$\,\kms\ \citep{She11}. Seven out of 12 (58\,\%) of the objects in our sample that are not double-peaked emitters and have fits to both \ion{Mg}{2} and \ion{C}{4} show \ion{C}{4} to \ion{Mg}{2} blueshifts exceeding the 1-$\sigma$ limits of the SDSS sample distribution ($>1760$\,\kms). This fraction for the SDSS comparison sample is only 107/728 (15\,\%). In Figure 7 top panels, we show the best fit broad line models of the five objects in our sample with the largest \ion{C}{4} to Balmer blueshifts. The H$\beta$, H$\alpha$, \ion{Mg}{2}, and \ion{C}{4} fits are plotted on top of each other, normalized in height and shown relative to the H$\alpha$ redshift. Interestingly, the spectra showing the largest \ion{C}{4} blueshifts ($\sim$5000--6000\,\kms, J0946+28, J1522+52) have sequentially decreasing, but measurable blueshifts toward \ion{Mg}{2} and H$\beta$. The blueshifted \ion{C}{4} profiles often appear asymmetric, skewed towards extreme blueshifts ($\sim$10,000\,\kms), and the asymmetry continues to appear in some of the \ion{Mg}{2} and Balmer lines. 

\begin{figure*}
\centering
\includegraphics[scale=.99]{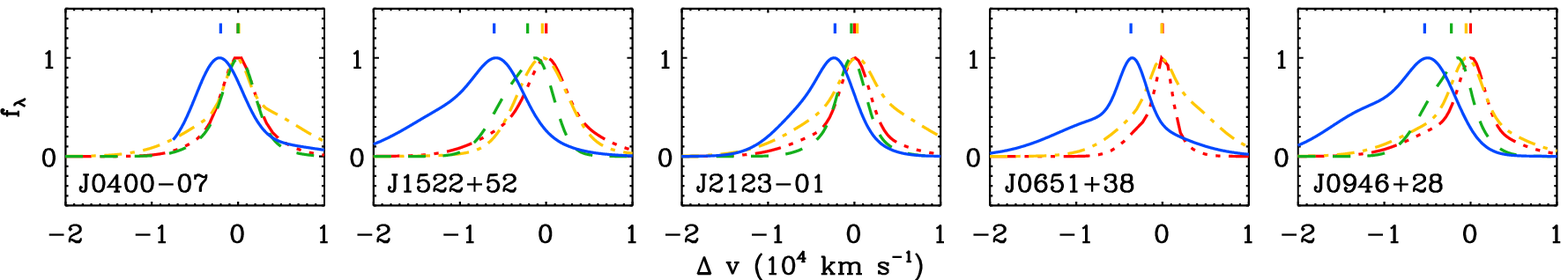}\vspace{5pt}
\includegraphics[scale=.99]{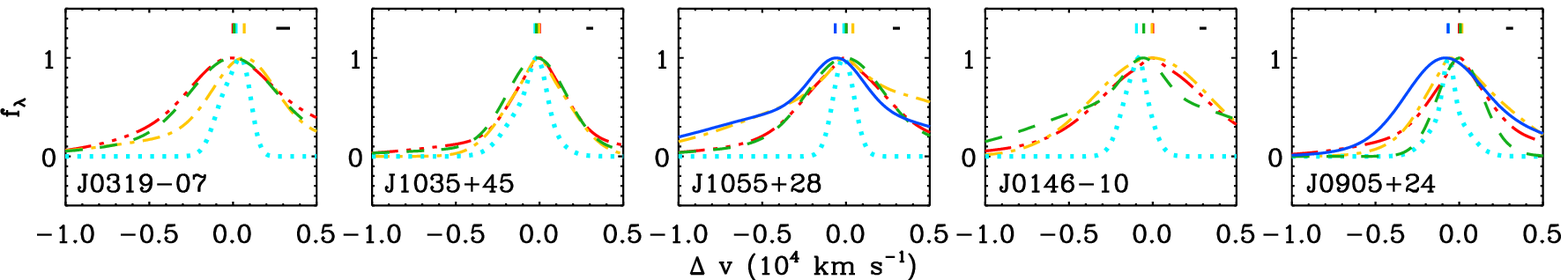}
\caption{Top: The normalized, broad line model profiles of H$\alpha$ (red triple dot-dashed), H$\beta$ (yellow dot-dashed), \ion{Mg}{2} (green dashed), and \ion{C}{4} (blue solid) for five objects with the most highly blueshifted \ion{C}{4} emission, along the line of sight velocity measured with respect to the broad H$\alpha$ redshift. Negative velocities indicate blueshift, and the centers of each broad line model are marked. The colors for the lines follow those of Figure 3. Bottom: The normalized, [\ion{O}{3}], H$\alpha$, H$\beta$, \ion{Mg}{2}, and \ion{C}{4} model profiles for five objects with [\ion{O}{3}] $\rm FWHM>1000$\,\kms\ at peak S/N$>$5, along the line of sight velocity measured with respect to the broad H$\alpha$ redshift. The [\ion{O}{3}] are colored cyan and dotted, while other lines are visualized as the top panels. The instrumental resolution elements are denoted as black horizontal lines on the top right of each panel.} 
\end{figure*}

We follow \citet{Bas05} to place these blueshifted \ion{C}{4} quasars along the shape, asymmetry, and shift parameters describing its profile. The shape parameter ranges within $1.05$--$1.46$ (1.16 on average) for the 11 objects with \ion{C}{4} blueshift smaller than 2000\,\kms, and $1.05$--$1.49$ (1.16 on average) for the five objects with \ion{C}{4} blueshift larger than 2000\,\kms. The indistinguishable distribution of the shape parameter indicates that the FWHM is a good indicator of the overall line shape, irrespective of the \ion{C}{4} blueshift (but see also, \citealt{Coa16} for the changing ratios between FWHM and $\sigma$ along \ion{C}{4} blueshift). On the other hand, the asymmetry parameter is preferentially distributed towards excess blue wings at highly blueshifted \ion{C}{4}, $-0.10$--$0.12$ ($0.02$ on average) and $-0.04$--$0.35$ (0.20 on average) for objects with \ion{C}{4} blueshift smaller and larger than 2000\,\kms, respectively. Furthermore, the shift parameter goes more negative at highly blueshifted \ion{C}{4}, $-0.32$--$-0.10$ ($-0.21$ on average) and $-0.67$--$-0.31$ ($-0.41$ on average) for objects with \ion{C}{4} blueshift smaller and larger than 2000\,\kms, respectively. Having seen the asymmetric, blueshifted nature of the \ion{C}{4} profiles that are suggestive of obscuration or outflows in \citet{Bas05}, we investigate if the \ion{C}{4} $M_{\rm BH}$ shows any systematic offset to the Balmer $M_{\rm BH}$ at higher blueshift. Indeed, we find that the \ion{C}{4} $M_{\rm BH}$ values of the five objects showing \ion{C}{4} to H$\alpha$ blueshifts $>$2000\,\kms\ (J0400-07, J0651+38, J0946+28, J1522+52, J2123-01), are positively offset with respect to the H$\alpha$ $M_{\rm BH}$ values by 0.26--0.68 (0.41 on average) dex. 

\begin{figure*}
\centering
\includegraphics[scale=.95]{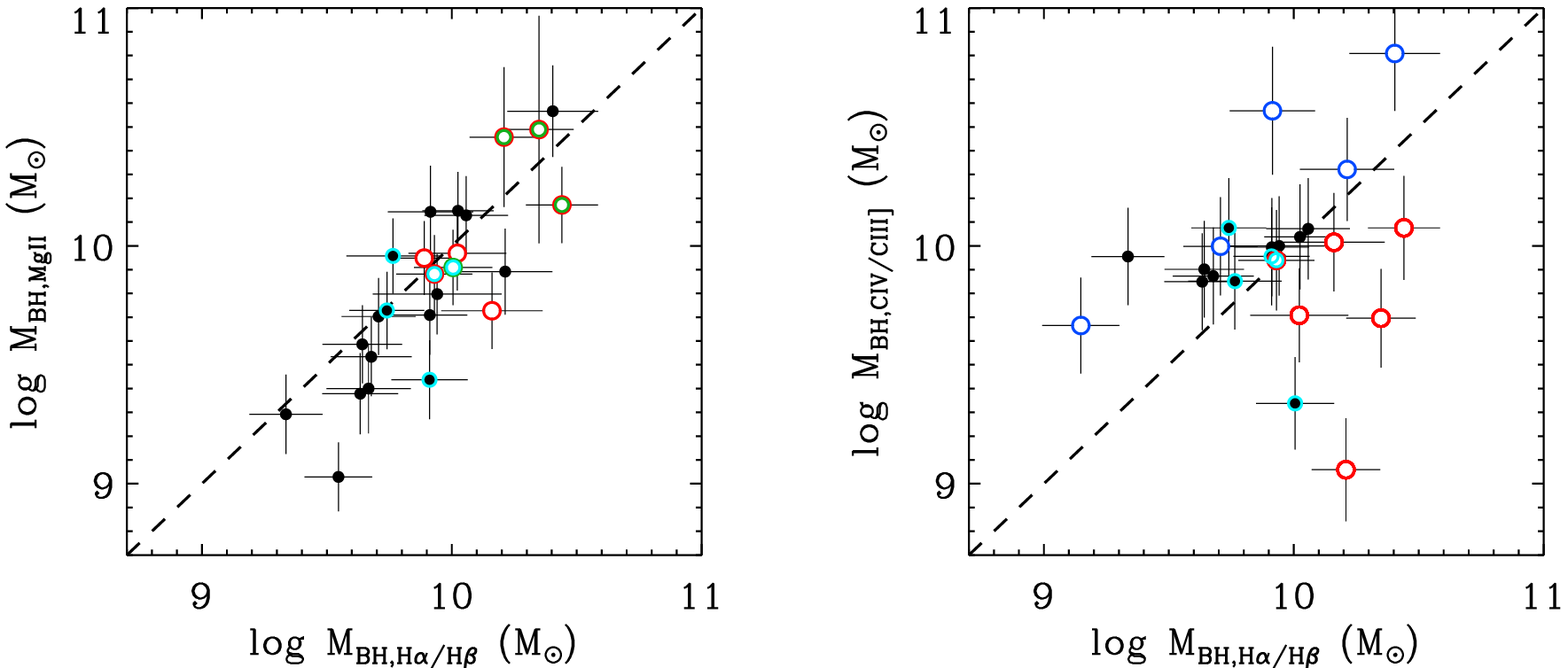}
\caption{Comparison of rest-optical and rest-UV $M_{\rm BH}$ values at the massive end. The H$\alpha$ and H$\beta$ $M_{\rm BH}$ values are averaged when $\rm S/N_{\rm H\beta}>10$; H$\alpha$ values are used otherwise. The \ion{C}{3}] $M_{\rm BH}$ values are plotted when \ion{C}{4} is absent; \ion{C}{4} values are used otherwise. Sources with double-peaked H$\alpha$ emission (red), extremely broad \ion{Fe}{2} around \ion{Mg}{2} (green), and broad \ion{C}{4} blueshifted by more than 2000\,\kms\ relative to broad H$\alpha$ (blue), have significant effect on the mass estimates and are colored with open circles. [\ion{O}{3}] lines with $\rm FWHM>1000$\,\kms\ (cyan) are negligibly ($\lesssim0.1$\,dex) affecting the mass estimates and are colored with filled circles.} 
\end{figure*}

\subsubsection{Ionized Outflows}
Last, we find a handful of blueshifted narrow [\ion{O}{3}]$\lambda$5007 emission that are wider than the typical narrow lines. Nine out of the 21 objects with a H$\beta$ region fit have unambiguous [\ion{O}{3}] profiles and peak S/N$>$5, where five of them meet $\rm FWHM_{[O\,III]}>1000$\,\kms\ (J0146-10, J0319-07, J0905+24, J1035+45, J1055+28). In Figure 7 bottom panels we plot the best fit model for the [\ion{O}{3}] and broad emission lines of these five objects. We find that the [\ion{O}{3}] profiles are typically blueshifted by a few hundred \kms\ relative to the broad Balmer redshift, and the FWHMs reach up to 1600--1900\,\kms\ (J0146-10, J0319-07, J1035+45). These [\ion{O}{3}] line widths are too broad to be explained by even the most massive galaxy's gravitational potential ($\rm FWHM\sim1000$\,\kms), and are broad relative to quasars at comparable luminosity or redshift (e.g., \citealt{Net04}; \citealt{Bru15}; \citealt{She16}). Previous work on broad [\ion{O}{3}] emission in quasars shows that the width correlates with its blueshift, indicative of strong outflows (e.g., \citealt{Liu14}; \citealt{Zak14}). We therefore investigate the effect of fixing the width of narrow lines around the H$\alpha$ for the objects with [\ion{O}{3}] profiles broader than $\rm FWHM=1000$\,\kms, bearing in mind they were fixed to 1000\,\kms\ in section 3. We fit the H$\alpha$ region by first fixing the narrow line FWHM to that of the [\ion{O}{3}] assuming all the narrow lines are fully broadened as the [\ion{O}{3}], and to 400\,\kms\ (the mean $\rm FWHM_{H\alpha}$ of local quasars used in section 3) assuming they are completely absent of outflows, respectively, where the $\rm FWHM_{H\beta}$ of quasars with broad [\ion{O}{3}] for example, seem to lie in between \citep{Zak14}. When the narrow line widths are fixed to that of the [\ion{O}{3}] instead of 1000\,\kms\ the H$\alpha$ $M_{\rm BH}$ values vary by $-0.01$ to $0.12$ (0.03 on average) dex, and by $-0.07$ to $0.02$ ($-0.02$ on average) dex when fixed to 400\,\kms. The limited differences in the $M_{\rm BH}$ values indicate that the effect of narrow line outflows, whether or not present at the H$\alpha$ region, is negligible in determining the broad line widths of extremely massive quasars. 

\subsubsection{Summary of biases from spectra}
In Figure 8, we compare the rest-optical and rest-UV $M_{\rm BH}$ values determined from section 3, marking the sources with unusual spectral features dealt here. We calculate how much the $\sigma_{\rm int}$ values between the rest-optical and rest-UV $M_{\rm BH}$ values decrease as we remove each class of unusual spectra. There is a general agreement between the Balmer and rest-UV line based masses up to $\sim$\,10$^{10}M_{\odot}$ with a much smaller scatter between the Balmer and \ion{Mg}{2} based masses ($\sigma_{\rm int}<0$) than between the Balmer and \ion{C}{4} based masses ($\sigma_{\rm int}=0.33$\,dex). This is in accord with earlier results from relatively less massive regimes (e.g., \citealt{She08}; J15). The $\sigma_{\rm int}$ between the Balmer and \ion{C}{4} $M_{\rm BH}$ values drops from $\sigma_{\rm int}=0.33$\,dex to 0.23\,dex when the double-peaked emitters are excluded, and down to $\sigma_{\rm int}=0.16$\,dex when objects with \ion{C}{4} blueshifts $>$2000\,\kms\ are further omitted. The number of $M_{\rm BH}>$\,10$^{10}M_{\odot}$ AGNs drops from 10, 7, 8 based on Balmer, \ion{Mg}{2}, and \ion{C}{4} based measurements, to 5, 4, 5 after removing the double-peaked H$\alpha$ emitters, extremely broad \ion{Fe}{2} around the \ion{Mg}{2}, and highly blueshifted \ion{C}{4} sources, respectively. This suggests that $M_{\rm BH}$ values $\gtrsim$\,10$^{10}M_{\odot}$ from any line should be carefully inspected for unusual features appearing in, or on top of the broad lines.

\subsection{Mass biasing factors in using the estimator}
\subsubsection{$f$-factor}
When bringing the spectral measurements into the single-epoch $M_{\rm BH}$ estimators, we consider the variations in the constant of the $M_{\rm BH}$ equation for AGNs ($f$--factor) that gives an overall normalization but is inaccurate for individual mass measurements. Because this constant is obtained from normalizing the zeropoint of the $M_{\rm BH}$--$\sigma_{*}$ relation, it has a systematic uncertainty of 0.3--0.4\,dex (e.g., \citealt{Kor13}; \citealt{McC13}). Using a constant $f$--factor as a representative value could overestimate the $M_{\rm BH}$ values for objects with anisotropic radiation or velocity dispersion (e.g., \citealt{Pet04}), when the line of sight values of these quantities are observed to be larger than geometrically averaged. To check whether the $M_{\rm BH} \sim 10^{10}M_{\odot}$ estimates can be explained by large line of sight spectral quantities of less massive BHs, we compared the average and rms scatter of the $L_{5100}$ and FWHM$_{\rm{H}\alpha}$ from our sample excluding the double-peaked emitters, divided by groups with $M_{\rm BH, H\alpha}$ values smaller and larger than the median, 10$^{9.9}M_{\odot}$. The averaged luminosities are $\log (L_{5100}/\ergs)=46.38 \pm 0.32$ and $46.55 \pm 0.38$ respectively, where the average difference in the luminosities correspond to a 0.08\,dex difference in $M_{\rm BH}$, much smaller than the difference in the average $M_{\rm BH}$ between the two groups, 0.52\,dex. This suggests that EMBH masses are not caused by the continuum luminosities that are boosted to unusually large values due to mechanisms like anisotropic accretion or gravitational lensing. Meanwhile, the averaged line widths are FWHM$_{\rm{H}\alpha}=3920 \pm 960$ and $6210 \pm 1390$\kms, respectively, showing that $\gtrsim$\,10$^{10}M_{\odot}$ estimates are influenced by large FWHM values that could be caused by anisotropic velocity field. However, this does not rule out the case where the line widths of extremely massive AGNs are intrinsically wide due to the stronger gravitational potential from the BH. 

There are issues of whether the $f$--factor is systematically different (up to $\sim 0.3$\,dex) between AGN subsamples grouped by host galaxy and BH properties, and also the limited statistical significance and dynamic range in the constraints to the $f$--factor (e.g., \citealt{McC13}; \citealt{Woo13}; \citealt{Ho14}). The local $M_{\rm BH}$--$\sigma_{*}$ relation for AGNs at least, which covers up to $\sim10^{9}M_{\odot}$ BHs, does not differ in $\sigma_{\rm int}$ with respect to inactive galaxies (e.g., \citealt{Woo13}). The spatially resolved direct dynamical $M_{\rm BH}$ measurement for inactive galaxies is thought to be much more accurate than the $M_{\rm BH}$ estimate for AGNs using a constant $f$--factor, and the $\sigma_{\rm int}$ for AGNs is expected to be significantly larger than that for inactive galaxies if there was a large intrinsic dispersion in the $f$--factor. The indistinguishable $\sigma_{\rm int}$ values for AGNs support the $M_{\rm BH}$--$\sigma_{*}$ relation itself is intrinsically scattered rather than the $f$--factor, which hints that the EMBHs in luminous quasars are intrinsically massive rather than positively biased in mass. Alternatively, the widespread distribution of $\rm FWHM_{H\beta}$ to \ion{Fe}{2} strengths for type-1 quasars are interpreted as the geometric orientation playing a significant role in the observed dispersion of the $\rm FWHM_{H\beta}$ values (e.g., \citealt{She14}). Still, the luminous, intermediate redshift type-1 quasar samples of S12 and \citet{She16} reaching up to EMBH masses show broader $\rm FWHM_{H\beta}$ values than the less luminous, local quasars in \citet{She11}, which they claim as due to intrinsically broader line width or more massive BHs for the S12, \citet{She16} samples. Further study of the $M_{\rm BH}$--$\sigma_{*}$ relation at the massive end, especially for active galaxies, and detailed modeling of the velocity structure of the BLR (e.g., \citealt{Bre11}; \citealt{Pan14}) are crucially required to better understand whether EMBH masses are either a geometric selection or intrinsic property.

\begin{figure}
\centering
\includegraphics[scale=.9]{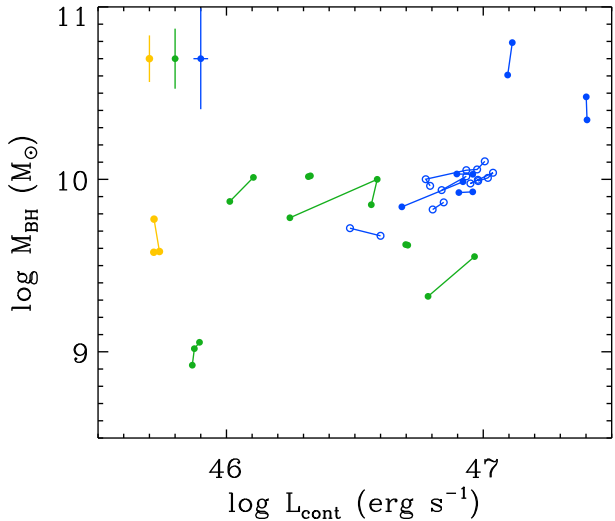}
\caption{The single-epoch $M_{\rm BH}$ values for objects with multiple epoch SDSS spectra, plotted against the continuum luminosity $L_{\rm cont}$ ($L_{5100}$ for H$\beta$, $L_{3000}$ for \ion{Mg}{2}, $L_{1350}$/$L_{1450}$ for \ion{C}{4}). The colors for the H$\beta$, \ion{Mg}{2}, and \ion{C}{4} line based values follow those in Figure 2. $M_{\rm BH}$ values from the same object are connected, and we mark the \ion{C}{4} masses open and filled circles to differentiate the objects. The mean $M_{\rm BH}$ uncertainty from each line based estimators, are shown in the upper left.} 
\end{figure}

\subsubsection{Variability}
Second, the $M_{\rm BH}$ estimators could suffer from variability such that the single-epoch measurements may not be representative values. Also, the time lag of the continuum to reach the BLR hinders obtaining coherent continuum luminosity and broad emission width from a given epoch. To probe the extent of continuum variability, we compiled the Catalina Real-time Transient Survey (CRTS; \citealt{Dra09}) optical light curves of our sample spanning 8 years on average. We calculated the variability amplitude $\sigma_{\rm var}$ where $\sigma_{\rm var}^{2}=\Sigma_{i=1}^{N}\{(m_{i}-\bar{m})^{2}-\Delta m_{i}^{2})\}/(N-1)$ for the mean magnitude $\bar{m}$ and $N$ magnitude and error measurements ($m_{i},\Delta m_{i}$), and find the $\sigma_{\rm var}$ value to range within 0.27 (median of 0.08) magnitudes. This level of intrinsic variation in the optical continuua is small, and even the object with the largest magnitude variation has a corresponding luminosity variation of 0.11\,dex, or a $M_{\rm BH}$ variation of 0.05\,dex. To further investigate the emission line variability, we plot in Figure 9 the single-epoch $M_{\rm BH}$ values from multi-epoch SDSS spectroscopy with connected symbols. We do not perform secondary flux calibration to the spectra (e.g., section 2.2) so that the variations in the spectral continuum includes the contribution from imperfect spectral flux calibration. Excluding the single object without a converging \ion{Fe}{2} solution around the \ion{Mg}{2} region (J0146-10), we have 1, 6, 10 objects with multi-epoch (2--5 visits) $M_{\rm BH}$ measurements in H$\beta$, \ion{Mg}{2}, and \ion{C}{4}, respectively. We find that the $M_{\rm BH}$ values fall within their errors throughout the sparsely covered epochs. Overall, the minor level and effect of variability on the single-epoch $M_{\rm BH}$ estimates for extremely massive AGNs, is consistent with the trends at lower masses (e.g., \citealt{Par12}; \citealt{Jun13}). 

\subsubsection{Overestimated ionizing continuum}
Third, we investigate cases where the observed AGN luminosities and broad line widths may not be applied to the standard $M_{\rm BH}$ equation. J15 report that the rest-optical continuum luminosity of extremely luminous AGNs ($L_{5100} \sim 10^{47}$\ergs) marginally overestimates the ionizing luminosity as traced by the H$\alpha$ line luminosity, perhaps hinting that the accretion disk of extremely massive and low spin BHs does not produce sufficient ionizing radiation (e.g., \citealt{Lao11}; \citealt{Wan14}). In Figure 10 we examine the luminous end $L_{5100}$--$L_{\rm{H}\alpha}$ relation, including our IRTF data points. We find that the IRTF data are mildly below the J15 relation, but does not show a systematic trend with $L_{5100}$. Further imposing a 20\% uncertainty limit to the combined data, most of the negatively offset outliers from J15 are removed so that the downward trend of the relation at the highest luminosities is less likely with higher sensitivity data. We also find that the IRTF data improves the completeness of the relation at $L_{5100} \sim 10^{46}$\ergs, filling the weaker emission line AGNs less covered by S12. The combined, sensitivity cut data in Figure 10 show a $0.03 \pm 0.13$\,dex offset and scatter to the J15 relation, supporting that the slope of the $L_{5100}$--$L_{\rm{H}\alpha}$ relation stays universal across $L_{5100} \sim 10^{42-47}$\ergs\ and that cold accretion disks in low spin EMBHs, if any, have a minor effect ($\lesssim 0.1$\,dex) in positively biasing the $M_{\rm BH}$ estimates derived using $L_{5100}$ instead of $L_{\rm{H}\alpha}$. 

\begin{figure}
\centering
\includegraphics[scale=.9]{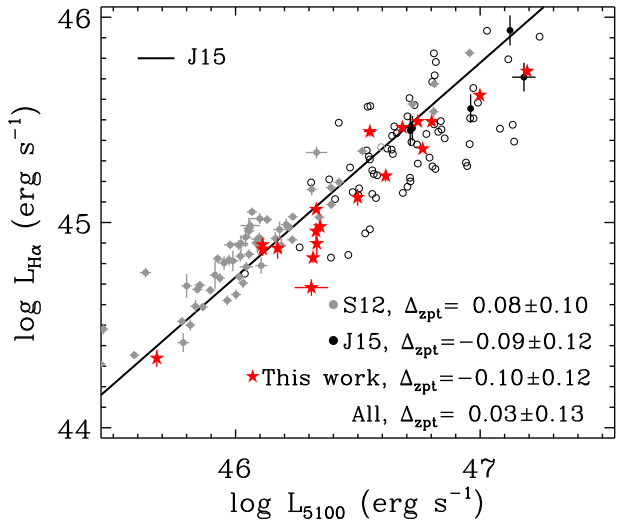}
\caption{The $L_{5100}$--$L_{\rm{H}\alpha}$ relation revisited at the luminous end. We plot data points from S12, J15, and this work (removing the double-peaked emitters) in gray, black, and red, and labeled S12, J15, and this work respectively. The luminosities are limited to having within 20\% uncertainty (filled symbols) and above 20\% uncertainty (open symbols). The J15 relation is overplotted as a black line, and the offset and 1-$\sigma$ scatter of the data points (within 20\% uncertainty) to the relation are listed for each sample.} 
\end{figure} 

\subsubsection{FWHM vs $\sigma$}
Last, we further look into the possible bias of using the broad line FWHM rather than the $\sigma$. Although the FWHM is technically simple to measure and is less affected than $\sigma$ by weakly constrained wings at poor sensitivity, it could be relatively inaccurate when the line profiles are far from a universal shape (e.g., \citealt{Pet04}; \citealt{Col06}). The $M_{\rm BH}$ estimators in J15 assume a constant $\rm{FWHM}=2\,\sigma$ condition for the broad H$\beta$ line widths, but any deviation from this constant could bias the $M_{\rm BH}$ estimates derived using FWHM. We checked if our sample exhibits this constant relation between the H$\alpha$ FWHM and $\sigma$ values determined from section 3.  We find that the mean and rms scatter of the FWHM to $\sigma$ ratios of the 21 objects without double-peaked emitters are mildly smaller than 2 ($\rm{FWHM}/\sigma=1.5\pm0.3$) and thus the EMBH masses are less likely to be spuriously overestimated by using FWHM instead of $\sigma$.
In fact, when assuming that the $M_{\rm BH, H\beta}$ scales proportional to $\sigma_{\rm H\beta}^{2}$ and $\rm{FWHM}_{\rm H\alpha}=2\,\sigma_{\rm H\alpha}$, the $\sigma_{\rm H\alpha}$ based $M_{\rm BH, H\alpha}$ values from Equations 1--2 would change from the FWHM based by $-0.09$--$0.55$ (0.30 on average) dex, nearly doubling the non-double-peaked, $M_{\rm BH, H\alpha}>10^{10}M_{\odot}$ objects from 7 to 13.
Interestingly, the FWHM to $\sigma$ ratios for the double-peaked emitters are somewhat larger than the rest of the sample (average and rms scatter of $2.2\pm0.4$) so that they will be better noticed by extremely wide FWHMs rather than $\sigma$ values.

\section{Discussion}
\subsection{$M_{\rm BH}$ bias due to double-peaked lines}
In the previous section, we considered cases where the $M_{\rm BH}$ estimates of AGNs including those in the EMBH regime, could be systematically biased. Here, we investigate if the two largest factors associated with possibly overestimated $M_{\rm BH}$ values, double-peaked broad emission and blueshifted \ion{C}{4}, are preferentially selected towards EMBH masses, or if they are conditionally appearing in general type-1 quasar spectra. We begin by comparing the double-peaked emitter fraction to those from the literature with larger samples at $z\lesssim0.4$. Our double-peaked emitter fraction, (5--7)/26 (19--27\%)\footnote{We consider J1010+05 and J1055+28 marginally double-peaked and provide the range of fractions depending on the inclusion of these objects.}, is comparable to or higher than 20\% among 106 radio-loud AGNs \citep{Era03}, and much higher than 3\% out of 3216 optically selected quasars \citep{Str03}, although the fraction is dependent on the parameter space where the double-peaked emitters are examined and the definition of being double-peaked. The double-peaked emitters show broader $\rm FWHM_{H\alpha}$ than typical AGNs, distributed mostly above 5000\,\kms\ and comparable in number to the non-double-peaked at above 8000\,\kms\ (\citealt{Era03}; \citealt{Str03}). Our study is in agreement with the expectations that 5/7 double-peaked emitters reach $\rm FWHM_{H\alpha}>\,8000$\,\kms\ while none of the rest of the objects' widths exceed this limit. 

We further examine if the double-peaked emitters generally have extremely wide FWHMs by using the visually classified double-peaked emitters in \citet{She11}. We cut their sample to $z<0.37$, S/N$>$10 to probe the double-peaked H$\alpha$ fraction, with their special interest flag selected as either highly double-peaked only, or highly/weakly double-peaked. Table 4 shows double-peaked emitter fractions per luminosity and FWHM bin, where the average uncertainty of the fractions are 0.34 and 0.21 times the fraction, for highly double-peaked cases and highly/weakly double-peaked cases respectively. We find that there is a mild increase of the double-peaked emitter fraction at higher $L_{5100}$ with a fixed FWHM, but the fraction increases more significantly with FWHM at a fixed $L_{5100}$. This suggests that extremely wide FWHMs are likely to be associated with double-peaked emitters, regardless of the $M_{\rm BH}$. We note that some double-peaked emitters could be missed for a variety of reasons. For instance, the line-emitting accretion disk model (e.g., \citealt{Che89}) predicts that the double-peaks may not be detached at small inclination angles ($i\lesssim10\deg$) and look alike ordinary broad emission. This adds ambiguity of whether the observed broad lines in type-1 AGNs are coming from random motions of broad line clouds or Keplerian rotation of a disk, and it may be separated by velocity resolved reverberation measurements of the line emitting region size (e.g., \citealt{Die98}; \citealt{OBr98}).

\renewcommand{\tabcolsep}{2.5pt}
\begin{deluxetable}{c|ccccc}
\tablecolumns{6}
\tabletypesize{\scriptsize}
\tablecaption{Double-peaked emitter fraction along FWHM and $L_{5100}$}
\tablewidth{0.45\textwidth}
\tablehead{
\colhead{} & \colhead{} & \colhead{} & \colhead{$\rm FWHM_{H\alpha}$} & \colhead{} & \colhead{}\\
\colhead{$\log L_{5100}$} \vline& \colhead{$<$2000} & \colhead{2000--4000} & \colhead{4000--6000} & \colhead{6000-8000} & \colhead{$>$8000}}
\startdata
44.6--44.9 &0.00--0.02 & 0.02--0.12 & 0.14--0.36 & 0.22--0.59 & 0.78--0.83\\
44.3--44.6 &0.00--0.00 & 0.01--0.13 & 0.12--0.31 & 0.14--0.52 & 0.23--0.48\\
44.0--44.3 &0.00--0.00 & 0.02--0.12 & 0.05--0.23 & 0.17--0.38 & 0.27--0.44
\enddata
\tablecomments{The fraction of $z<0.37$, S/N$>10$, type-1 quasars in \citet{She11} that are classified as highly double-peaked, and highly/weakly double-peaked are shown in ranged values. The $L_{5100}$ and $\rm FWHM_{H\alpha}$ are in units of \ergs\ and \kms, respectively.} 
\end{deluxetable} 

\begin{figure*}
\centering
\includegraphics[scale=.665]{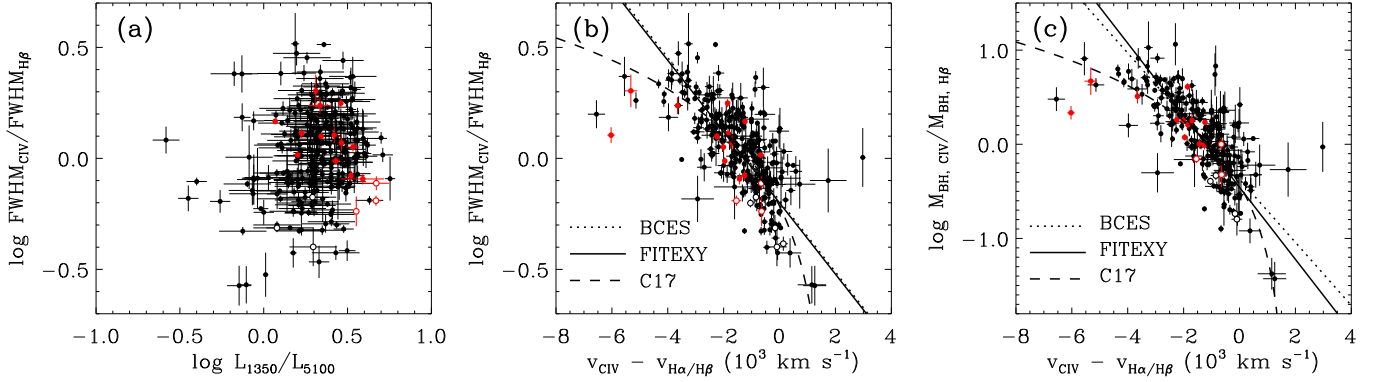}
\caption{The ratio between \ion{C}{4} and Balmer broad line FWHMs plotted against rest-frame 1350--5100\,\AA\ continuum luminosity ratio (left) and Balmer to \ion{C}{4} broad line shift (center), and the $M_{\rm BH}$ ratio against \ion{C}{4} blueshift (right). Negative velocities indicate blueshift. The Balmer to \ion{C}{4} shifts are averaged when both the H$\alpha$ and H$\beta$ lines are observed, and the FWHM$_{\rm H\beta}$ is converted from FWHM$_{\rm H\alpha}$ whenever available using the J15 relation, in order to benefit from the enhanced sensitivity of the H$\alpha$ line. We plot the data points from the literature (\citealt{Net07}; \citealt{Sha07}; \citealt{Die09}; \citealt{Ass11}; \citealt{Ho12}; S12; \citealt{Ben13}; \citealt{Par13}; J15; \citealt{Mej16}; C17 and the references within) converted to our adopted cosmology in black, and from this work in red. We limit the line width ratios and luminosity ratios to sources with $<0.15$\,dex uncertainties, and the line shifts to sources with $<1000\kms$ uncertainties. Objects with broad absorption lines near the \ion{C}{4} line are rejected, while objects with $\rm FWHM_{H\alpha}>8000\kms$ among the literature or those classified as double-peaked emitters from this work are shown in open circles. The two objects in \citet{Sha07} that overlap with \citet{Par13} are removed, while the $L_{1350}$ and $\rm FWHM_{C_{\,IV}}$ from \citet{Net07} are updated to the values from \citet{She11}. On panel (b) the FITEXY and BCES fit to the filled data are shown in solid and dotted lines, respectively, while the C17 relation (corrected as $M_{\rm BH} \propto \rm FWHM^{2}$ for panel (c)) is overplotted (dashed line).}
\end{figure*} 

Interestingly, the \ion{Mg}{2} spectra of double-peaked H$\alpha$ emitters also often exhibit double peaks that are weaker or appear blended (e.g., \citealt{Era03}; \citealt{Era04}; \citealt{Era15}). These features in our sample are weak (J0203+13) or hard to tell (J0257+00, J0741+32, J1010+05, J1053+34, J1055+28, J1057+31), somewhat consistent with the literature, and explains why our double-peaked H$\alpha$ emitters were not flagged out by rest-UV spectra in section 2.1. This indicates the likelihood that double-peaked emission are ambiguously mixed on top of the broad \ion{Mg}{2} line, placing negative implications on the reliability of $M_{\rm BH}$ measurement from the \ion{Mg}{2} line alone at wide FWHM values. Furthermore, we reviewed that extremely wide \ion{Fe}{2} around \ion{Mg}{2} could be associated with overestimated \ion{Mg}{2} width solutions (section 4.2.2). The lower limit of FWHM$_{\rm Mg_{\,II}}$ when this occurred in our sample is 6800\kms\ from our analysis, and 6900\kms\ from the automated spectral fitting of \citet{She11}. Caveats of using the broadest $\rm FWHM_{\rm Mg_{\,II}}$ for $M_{\rm BH}$ measurements are in line with existing studies where the rotational broadening is able to fully explain the observed FWHMs only up to 4000--6500\,\kms\ in typical BLRs (e.g., \citealt{Kol13}; \citealt{Mar13}). We also note that the FWHMs of the highly blueshifted ($>$\,2000\,\kms\ relative to H$\alpha$, section 4.2.3) \ion{C}{4} profiles are 5400--11100\,\kms\ (8200\,\kms\ on average), near the broad end of the FWHM distribution. 

To summarize, we caution against blindly adopting $M_{\rm BH}$ estimates based on any line with $\rm FWHM\gtrsim$\,8000\,\kms.
For example, searching for quasars in \citet{She11} with S/N$>$10 and flagged not to be double-peaked emitters, there are 14 H$\alpha$-based and 213 H$\beta$-based masses with $M_{\rm BH}>10^{9.5}M_{\odot}$ at $z<0.37$ and $z<0.7$, respectively. However, 14/14 H$\alpha$-based and 212/213 H$\beta$-based objects have broad line FWHM\,$>$\,8000\,\kms\ and the spectra of these objects need to be carefully checked. Indeed through visual inspection of the spectra, we find that 12 out of the 14 H$\alpha$ spectra indicating $M_{\rm BH}>10^{9.5}M_{\odot}$ and FWHM\,$>$\,8000\,\kms\ show moderate to strong double-peaked line profiles, giving cautions about their $M_{\rm BH}$ values.

\subsection{$M_{\rm BH}$ bias due to \ion{C}{4} blueshift}
Next, we consider the general effect of blueshifted \ion{C}{4} on $M_{\rm BH}$ estimation, thought to be a combined effect of outflows and obscuration in high ionization lines (e.g., \citealt{Bas05}). Though it is expected that optically bright type-1 AGNs are seen through minimal obscuring material, they display a moderate range of UV/optical through infrared colors (e.g., \citealt{Ric03}; \citealt{Jun13}). This hints that not only the UV continuum emission can be absorbed, but likewise for the broad line emission so that the reliability of UV line widths should be checked, especially at higher levels of obscuration. We follow S12 to plot in Figure 11(a) the ratio between the \ion{C}{4} and Balmer broad line FWHMs against rest-frame 1350--5100\,\AA\ continuum color, using compiled references and this work. We checked that the plotted objects are luminous enough to have an estimated host galaxy contamination of less than 20\% at 5100\,\AA\ \citep{She11}, or have Hubble Space Telescope imaging so that the spatially resolved host galaxy contamination is below 20\% at optical wavelengths. We do not find any correlation between the quantities (linear Pearson correlation coefficient $r$=0.18), implying the \ion{C}{4} line width does not suffer any more systematic biases than the Balmer lines. Instead, having checked that the IRTF sources with blueshifted \ion{C}{4} emission show broader \ion{C}{4} than the Balmer line widths (section 4.2), we plot in Figures 11(b)--(c) the \ion{C}{4} to Balmer broad line FWHM and $M_{\rm BH}$ ratios against the Balmer to \ion{C}{4} broad line shift from compiled references and from this work. We find that the quantities are positively correlated ($r$=0.66 and 0.70 respectively), in accord with the trends between the \ion{C}{4} and \ion{Mg}{2} (e.g., \citealt{She08}). The linear fit to the data based on the FITEXY and BCES methods (\citealt{Pre92}; \citealt{Akr96}) respectively yield
\begin{eqnarray}\begin{aligned}
\log (\rm FWHM_{C_{\,IV}}/FWHM_{H\beta})=-(0.201 \pm 0.003)\\
\rm -(0.161 \pm 0.002)(v_{C_{\,IV}}-v_{H\alpha/H\beta})\,(10^{3}\,km\,s^{-1})\\
\log (\rm FWHM_{C_{\,IV}}/FWHM_{H\beta})=-(0.190 \pm 0.015)\\
\rm -(0.161 \pm 0.010)(v_{C_{\,IV}}-v_{H\alpha/H\beta})\,(10^{3}\,km\,s^{-1}).
\end{aligned}\end{eqnarray}
\begin{eqnarray}\begin{aligned}
\log (\rm M_{BH, C_{\,IV}}/M_{BH, H\beta})=-(0.457 \pm 0.006)\\
\rm -(0.382 \pm 0.003)(v_{C_{\,IV}}-v_{H\alpha/H\beta})\,(10^{3}\,km\,s^{-1})\\
\log (\rm M_{BH, C_{\,IV}}/M_{BH, H\beta})=-(0.371 \pm 0.035)\\
\rm -(0.335 \pm 0.022)(v_{C_{\,IV}}-v_{H\alpha/H\beta})\,(10^{3}\,km\,s^{-1}).
\end{aligned}\end{eqnarray}
Because of the tighter linear correlation for the $M_{\rm BH}$ ratios than the FWHM ratios, we recommend using Equation (4) when correcting the \ion{C}{4} $M_{\rm BH}$ values. Equations (3)--(4) imply that \ion{C}{4} to Balmer FWHM ratios systematically increase with \ion{C}{4} blueshift (e.g., \citealt{Coa16}; \citealt{Coa17}, hereafter C17), for instance, by 0.32\,dex between 0 and 2000\,\kms\ \ion{C}{4} blueshift, or by 0.67--0.76\,dex in $M_{\rm BH, C_{\,IV}}/M_{\rm BH, H\beta}$ values.

The blueshift of the \ion{C}{4} line has been considered as one of the causes for the scatter in the broad line width ratios against \ion{Mg}{2} or Balmer lines. At the time of writing, we find the C17 relation well points out for the systematic overestimation in \ion{C}{4} to Balmer line width ratios along \ion{C}{4} blueshift, drawing similar conclusions although linear in correction method as opposed to our log-linear correction. We compare the reduction in the intrinsic scatter between the \ion{C}{4} to H$\beta$ $M_{\rm BH}$ ratios when applying either corrections to the data in Figure 11(c), for the \ion{C}{4} blueshift bounded within $-$1000 and 5000\kms\, in order to compare well sampled data and to reject data where the C17 relation diverges. We find that $\sigma_{\rm int}$ decreases merely from 0.38 to 0.27 (this work, FITEXY), 0.25 (this work, BCES), and 0.30 (C17 relation) dex. The C17 relation performs as much as ours (or perhaps better at \ion{C}{4} blueshifts larger than 5000\kms) to reduce the $\sigma_{\rm int}$ values, considering that we are correcting for the $M_{\rm BH}$ ratios while using the FWHM$^{2}$ ratios from C17, although the linear correlation coefficients are slightly larger between the \ion{C}{4} blueshift and log $M_{\rm BH}$ ratio ($r=0.70$) than against $M_{\rm BH}$ ratio ($r=0.66$). In any case, the relatively minor change in $\sigma_{\rm int}$ values (0.08--0.13 out of 0.38\,dex) imply that the broad line outflows, although effectively explaining the bias in the \ion{C}{4} to Balmer $M_{\rm BH}$ ratios, are not fully responsible for the scatter.

Among other mutually correlated observables (Eigenvector 1, \citealt{Bor92}) that scale with the broad line width ratios or the residuals of the ratios are the \ion{C}{4} luminosity, equivalent width of the \ion{C}{4} line, and shape parameters (e.g., \citealt{Bas05}; \citealt{Run13}), reducing the intrinsic scatter between the \ion{C}{4} and Balmer based $M_{\rm BH}$ values from 0.43--0.51\,dex by merely 0.10--0.13\,dex. Many Eigenvector 1 properties are correlated with the Eddington ratio, perhaps hinting that the \ion{C}{4} line width bias could be driven by a physical mechanism such as strong outflowing winds at high Eddington ratios, although the high Eddington ratio is a necessary rather than sufficient condition for \ion{C}{4} outflows \citep{Bas05}. We further note that studies reporting the reduction of the $\sigma_{\rm int}$ value between the \ion{C}{4} and Balmer line based $M_{\rm BH}$ values by adding an obscuration correction term or adopting a shallower scaling of the $\rm FWHM_{C_{\,IV}}$ term (e.g., \citealt{Ass11}; S12) are not as effective when the dynamic range and sampling of the parameter space are improved (e.g., Figure 11(a) in this work, J15). Overall, the intrinsic scatter between \ion{C}{4} to Balmer line based $M_{\rm BH}$ values ($\sim0.4$\,dex, J15) is not yet fully explained by either empirical or physical approaches, leaving the possibility that the \ion{C}{4} mass estimator is less reliable than Balmer- or \ion{Mg}{2}-based estimators due to more fundamental reasons, e.g., non-virialized or non-reverberating velocity structure within the \ion{C}{4} line region \citep{Den12}. 

\begin{figure*}
\centering
\includegraphics[scale=.75]{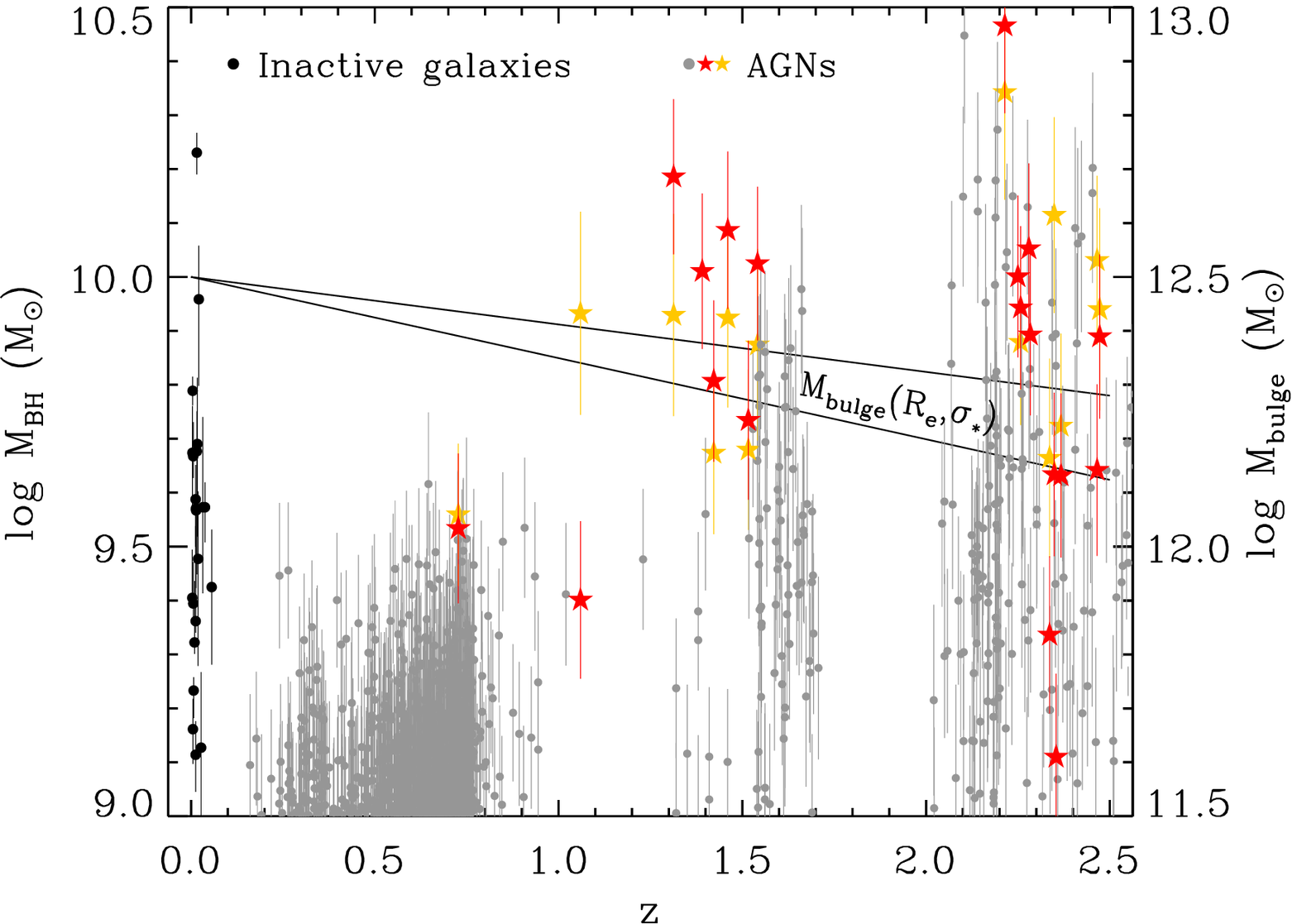}
\caption{The massive end of the black hole distribution as a function of redshift. For active galaxies Balmer $M_{\rm BH}$ values derived using the J15 estimator are plotted. The data come from this work (stars, H$\alpha$ in red and H$\beta$ in yellow) and from the literature (both H$\alpha$ and H$\beta$ in gray, \citealt{She04} \citealt{Die09}; \citealt{Ass11}; \citealt{She11}; S12; \citealt{Mat13}; C17 and the references within) converted to our adopted cosmology. Sources showing double-peaked broad emission or having broad FWHM\,$>$\,8000\,\kms\ are removed. For inactive galaxies in the local universe we plot the direct dynamical measurements (black, \citealt{Rus11}; \citealt{Kor13}; \citealt{Wal15}; \citealt{Tho16}; \citealt{Wal16}). We limit the data to sources with $M_{\rm BH}$ uncertainties less than 0.2\,dex, and with continuum S/N\,$>$\,10 for the SDSS DR9 objects (without measurement errors for the continuum luminosity and line width, thus their $M_{\rm BH}$ errors on the figure are underestimated). The model evolution of $M_{\rm bulge}$ assuming that it will become a fiducial 10$^{12.5}M_{\odot}$ bulge galaxy hosting a 10$^{10}M_{\odot}$ BH at $z=0$ (e.g., \citealt{McC13}) are overplotted for when using $M_{\rm bulge} \propto R_{e}\sigma_{*}^{2}$ (solid line), with the range of slopes obtained from the literature (section 5.2).} 
\end{figure*}

\subsection{EMBH-host galaxy coevolution}
We probe how the observed EMBH masses in AGNs constrain models for galaxy evolution in Figure 12, showing Balmer $M_{\rm BH}$ values at the massive end as a function of redshift from compiled references and from this work. The most massive BHs seen in local inactive galaxies appear similarly massive to AGNs at $z\sim1$ and beyond. Assuming that the EMBH hosting AGNs at $z=2.5$ will become inactive bulge galaxies falling on the local $M_{\rm BH}$--$M_{\rm bulge}$ relation, we draw in Figure 12 the expected evolutionary tracks of the $M_{\rm bulge}$ adopting the consensus of observed and simulated evolution of $\sigma_{*}$ and effective radius ($R_{e}$) for massive galaxies, i.e., 20--40\% decrease in $\sigma_{*}$ and 3--5 times increase in $R_{e}$ from $z=2$ to 0, and the proportionality $M_{\rm bulge} \propto R_{e}\sigma_{*}^{2}$ (e.g., \citealt{Tru06}; \citealt{Tof07}; \citealt{Cen09}; \citealt{Naa09}; \citealt{van10}; \citealt{Ose12}; \citealt{Tri16}). The estimated $M_{\rm bulge}$ values at $z\sim2$ are factors of few smaller than the most massive local galaxies, leaving the possibility for $M_{\rm bulge}$ to grow after the black hole has reached EMBH mass. Therefore, the observed EMBH masses in $z=1-2$ AGNs require them to be overmassive to their host bulges. 

On the other side, we consider the effect of relaxing the assumption that the most massive BHs are hosted in the most massive early-type galaxies shaping the present day BH--galaxy scaling relations. First, the host galaxies of $z=1-2$ EMBHs may not end up being the most massive galaxies due to galaxy environment. If the EMBH hosts are not the central galaxies in moderately dense environments (e.g., \citealt{Bro08}; \citealt{Wel16}), they will not encounter minor mergers as frequently, which would imply more limited size growth. Indeed, local examples of overmassive BHs are often in compact galaxies (e.g., \citealt{Rus11}; \citealt{van12}; \citealt{Fer15}; \citealt{Wal16}). Second, it could be that the host galaxy evolves to be massive in stellar content, but not bulge-dominated in morphology. Gas-rich major mergers that could trigger the observed AGN luminosity traced by our sample (e.g., \citealt{Hon15}) and form a bulge may still leave a disk, and transformation into the bulge through secular processes or repeated minor mergers could have somehow been prevented (e.g., \citealt{Spr05}; \citealt{Rob06}; \citealt{van11}). This scenario is consistent with most of the overmassive BHs on the $M_{\rm BH}$--$M_{\rm{bulge}}$ relation (e.g., \citealt{Wal16}) being lenticular galaxies, with bulge to total mass (or luminosity) ratios typically ranging below unity (0.1--0.6, e.g., \citealt{Cre99}; \citealt{Rus11}; \citealt{van12}; \citealt{Str13}; \citealt{Wal15}). 

We have discussed that the relative growth modes for extremely massive BHs and their host galaxies can not only depend on the galaxy mass, but also environment or morphology. Galaxy environment studies of EMBH hosts will help probe the contribution of mergers shaping the BH--galaxy scaling relations (e.g., \citealt{Jah11}), and spatially resolved imaging of the host will tell if a two parameter relation (e.g., $M_{\rm BH}$--$\sigma_{*}$) is sufficient to explain black hole-galaxy coevolution.  Luminous AGN activity is rare in the present day universe and EMBHs have mostly been found in quiescent early-type galaxies. Further discoveries of EMBHs (e.g., \citealt{van15}) in lower bulge masses will constrain how tight the BH--galaxy scaling relations are at their massive end, and how often EMBHs in distant AGNs remain in the most massive galaxies at present.

\section{Summary}
We performed followup rest-optical spectrocopy of a sample of 26 extremely massive quasars at $0.7<z<2.5$ in order to cross check their rest-UV $M_{\rm BH}$ values, and to examine possible biases affecting the measured $M_{\rm BH}$ values. We summarize the results as follows.\\

1. The rest-UV $M_{\rm BH}$ estimates of $\lesssim$10$^{10}M_{\odot}$ in luminous AGNs, are generally consistent with the Balmer based estimates. However, double-peaked emitters strongest in the H$\alpha$, extremely broad \ion{Fe}{2} around \ion{Mg}{2}, and highly blueshifted ($>$\,2000\,\kms) \ion{C}{4} profiles are frequently associated with $M_{\rm BH}\gtrsim$10$^{10}M_{\odot}$ estimates, easily boosting reported masses by a factor of a few. We find these cases mostly at broad line FWHM\,$>$\,8000\,\kms, and make cautionary remarks for estimating $M_{\rm BH}$ values based on any line width over this limit. The presence of broadened (FWHM\,$>$\,1000\,\kms) narrow emission (e.g., [\ion{O}{3}]), however, does not appear to significantly bias EMBH mass measurements.\\

2. We checked for systematic biases in single-epoch $M_{\rm BH}$ estimators for AGNs with EMBH masses and general AGNs. Anisotropic radiation and the use of broad line FWHM in place of $\sigma$ are not the major cause of producing false EMBH $M_{\rm BH}$ estimates for our sample. Furthermore, variability, overestimated line equivalent width from cold accretion disks, and obscuration do not bias the $M_{\rm BH}$ estimates for general type-1 quasars by more than $\sim$\,0.1\,dex. Instead, correcting the \ion{C}{4} $M_{\rm BH}$ estimator based on its blueshift relative to the Balmer line redshift, the \ion{C}{4} $M_{\rm BH}$ values decrease by 0.67--0.76\,dex from a zero to 2000\,\kms\ blueshift, with sizable scatter.\\

3. Removing the systematically uncertain $M_{\rm BH}$ values arising from the spectra or mass estimators, there is still a chance that EMBH masses are boosted by anisotropic motion of the broad line region from $\sim10^{9.5}M_{\odot}$ BHs, but this is contradictory to the current $\sigma_{\rm int}$ values of the local $M_{\rm BH}$--$\sigma_{*}$ relation for AGNs. The observed and simulated growth of $M_{\rm{bulge}}$ in massive galaxies support that EMBH hosting AGNs at $z=1-2$ are growing dominantly by minor dry mergers, with their BHs overmassive to the host's bulge mass. Depending on the galaxy environment in galactic and intergalactic scales, we expect that either the EMBH host will catch up the BH growth or the BH will stay overmassive to the bulge.

\acknowledgments
We thank the anonymous referee for the comments that greatly improved the paper, and the IRTF staff for their kind help during data acquisition at on-site and remote observations. 
This work was supported by the National Research Foundation of Korea (NRF) grant, No. 2008-0060544, funded by the Korea government (MSIP). This research was supported by an appointment to the NASA Postdoctoral Program at the Jet Propulsion Laboratory, administered by Universities Space Research Association under contract with NASA. 
H. D. J., M. I., and D. K. were visiting astronomers at the Infrared Telescope Facility, which is operated by the University of Hawaii under contract NNH14CK55B with the National Aeronautics and Space Administration.
This publication makes use of data products from the Two Micron All Sky Survey, 
which is a joint project of the University of Massachusetts and the Infrared Processing 
and Analysis Center/California Institute of Technology, funded by the National Aeronautics and 
Space Administration and the National Science Foundation.
This publication makes use of data products from the United Kingdom Infrared Deep Sky Survey. The United Kingdom Infrared Telescope (UKIRT) is supported by NASA and operated under an agreement among the University of Hawaii, the University of Arizona, and Lockheed Martin Advanced Technology Center; operations are enabled through the cooperation of the East Asian Observatory. When the data reported here were acquired, UKIRT was operated by the Joint Astronomy Centre on behalf of the Science and Technology Facilities Council of the U.K.
This publication makes use of data products from the Wide-field Infrared Survey Explorer, 
which is a joint project of the University of California, Los Angeles, and the Jet Propulsion 
Laboratory/California Institute of Technology, funded by the National Aeronautics and Space Administration.

\end{document}